\documentclass[a4paper,11pt]{article}

\usepackage[a4paper,left=2.73cm,right=2.7cm,top=3cm,bottom=3.5cm]{geometry}

\usepackage{amsmath,amssymb,graphicx,caption,subcaption,cite}

\usepackage[colorlinks=true,linktocpage=true,linkcolor=blue,citecolor=blue]{hyperref}

\addtocontents{toc}{\protect\setcounter{tocdepth}{2}}
\numberwithin{equation}{section}

\def\d{\mathrm{d}}
\newcommand{\mt}[1]{\textrm{\tiny #1}}
\newcommand{\Jkah}{J}
\newcommand{\etakah}{\eta}
\newcommand{\cN}{{\cal N}}
\newcommand{\cM}{{\cal M}}
\newcommand\sfh{\mathsf{h}}
\newcommand\sff{\mathsf{f}}
\newcommand\sfg{\mathsf{g}}
\newcommand\sfb{\mathsf{b}}
\newcommand\sfc{\mathsf{c}}
\newcommand{\lp}{\Lambda_\mt{LP}}
\newcommand{\rholp}{\varrho_\mt{LP}}
\newcommand{\bea}{\begin{eqnarray}}
\newcommand{\eea}{\end{eqnarray}}
\newcommand{\vr}{\varrho}
\newcommand{\dd}{\mathrm{d}}

\newcommand{\ls}{\ell_s}
\newcommand{\gym}{g_\mt{YM}}

\newcommand{\Qc}{Q_\mt{c}}
\newcommand{\Qf}{Q_\mt{f}}
\newcommand{\nc}{N_\mt{c}}
\newcommand{\nf}{N_\mt{f}}

\newcommand{\be}{\begin{equation}}
\newcommand{\ee}{\end{equation}}
\newcommand{\bal}{\begin{aligned}}
\newcommand{\eal}{\end{aligned}}

\newcommand{\eq}[1]{(\ref{#1})}
\newcommand{\eqq}[1]{Eq.~(\ref{#1})}
\newcommand{\Fig}[1]{Fig.~\ref{#1}}
\newcommand{\sac}{\, , \qquad}

\begin{document}

\begin{titlepage}

\thispagestyle{empty}

\begin{flushright}
\hfill{ICCUB-16-038}
\end{flushright}

\vspace{40pt}  
	 
\begin{center}

{\LARGE \textbf{Holography with a Landau pole}}
	\vspace{30pt}
		
{\large \bf Ant\'on F. Faedo,$^{1}$   David Mateos,$^{1,\,2}$   \\ [2mm]
Christiana Pantelidou$^{1}$  and Javier Tarr\'\i o$^{3}$}
		
\vspace{25pt}

{\normalsize  $^{1}$ Departament de F\'\i sica Qu\'antica i Astrof\'\i sica and Institut de Ci\`encies del Cosmos (ICC),\\  Universitat de Barcelona, Mart\'\i\  i Franqu\`es 1, ES-08028, Barcelona, Spain.}\\
\vspace{15pt}
{ $^{2}$Instituci\'o Catalana de Recerca i Estudis Avan\c cats (ICREA), \\ Passeig Llu\'\i s Companys 23, ES-08010, Barcelona, Spain.}\\
\vspace{15pt}
{ $^{3}$
Physique Th\'eorique et Math\'ematique, Universit\'e Libre de Bruxelles (ULB) \\
and International Solvay Institutes, Campus de la Plaine CP 231, B-1050, Brussels, Belgium.}

\vspace{40pt}
				
\abstract{
Holography for UV-incomplete gauge theories is  important but poorly understood. A paradigmatic example is $d=4$, 
$\mathcal{N}=4$ super Yang-Mills coupled to $\nf$ quark flavors, which possesses a Landau pole at a UV scale $\lp$. The dual gravity solution  exhibits a UV singularity at a finite proper distance along the holographic direction. Despite this, holographic renormalization can be fully implemented via analytic continuation to an AdS solution. The presence of a UV cut-off manifests itself in several interesting ways. 
 At energies $E \ll \lp$  no pathologies appear, as expected from effective field theory. In contrast, at scales $E \lesssim \lp$ the gravitational potential becomes repulsive, and at temperatures $T \lesssim \lp$ the specific heat becomes negative. Although we focus on $\mathcal{N}=4$ super Yang-Mills with flavor, our qualitative results apply to a much more general class of theories, since they only depend on the fact that the metric near the UV singularity is a hyper-scaling violating metric with exponent $\theta > d-1$.
}

\end{center}

\end{titlepage}

\tableofcontents

\hrulefill
\vspace{10pt}

%%%%%%%%%%%%%%%
%%%%%%%%%%%%%%%%%%%%%%%%%%%%%%%%
\section{Introduction}
%%%%%%%%%%%%%%%%%%%%%%%%%%%%%%%%
%%%%%%%%%%%%%%%
\label{intro}
 Quantum field theories with a Landau pole are ultraviolet (UV) incomplete but may nevertheless have interesting phenomenological applications. From the viewpoint of such applications one is usually interested in restricting the energy range to that sufficiently below the Landau pole, so that the results are fairly insensitive to the UV behavior of the theory. Our purpose in this paper is the opposite, namely to analyze from a holographic perspective the \emph{entire} RG flow of a gauge theory afflicted by a Landau pole. Our motivation is to shed light on holography for UV-incomplete theories in general, and to understand how holography encodes the physics of the Landau pole singularity in particular. 
 %DMDM
 In other words, rather than shielding ourselves from the Landau pole  by focusing on safe infrared (IR) physics, we are interested in understanding what UV features of the gauge theory, pathological or not, can be reliably studied with supergravity. 
 
The set of models that we will focus on consists of four-dimensional, quiver-like 
$\mathcal{N}=1$ super-conformal Yang-Mills theories coupled to $\nf$ massless hypermultiplets in the fundamental representation of the gauge group. This set  includes $\mathcal{N}=4$ super Yang-Mills (SYM) theory with gauge group 
SU($\nc$)  as a particular example. We will refer to the fundamental matter as ``flavor'' and, although it includes both bosonic and fermionic degrees of freedom, also as ``quarks''.  Since the $\beta$-function of the super-conformal theory without matter is exactly zero, the coupling to the quarks produces a positive $\beta$-function and a  Landau pole at a high energy scale $\lp$. 
 
The string dual of the gauge theory above is given by type IIB string theory on the background of $\nc$ ``color'' D3-branes and $\nf$ ``flavor'' D7-branes \cite{Karch:2002sh}.  If the compact space transverse to the D3-branes is S$^5$ then the gauge theory is $\mathcal{N}=4$ SYM; if it is any other Sasaki--Einstein manifold then the gauge theory is a quiver theory. The D3-D7 system has been extensively studied in the so-called ``probe approximation'' (see \cite{Kruczenski:2003be,Babington:2003vm} for early references and \cite{Erdmenger:2007cm,CasalderreySolana:2011us} for reviews) in which the backreaction of the D3-branes on spacetime is included but that of the D7-branes is neglected. In the gauge theory this corresponds to a quenched approximation in which the effect of the quarks on the dynamics of the gluons and the adjoint matter is ignored. This is justified over a large range of energies if $\nf/\nc \ll 1$. One way to see this is to note that the $\beta$-function is proportional to $\nf/\nc$ and hence the running of the coupling is a small correction to the physics over energy ranges that are not exponentially large in $\nc/\nf$. However, at sufficiently high energies the running of the coupling cannot be ignored and the probe approximation inevitably breaks down. To explore this high-energy regime  holographically one must go beyond the probe approximation and include the backreaction of the D7-branes. 

In order to do so we will follow the approach of \cite{Benini:2006hh} (see \cite{Nunez:2010sf} for a review) and smear the $\nf$ D7-branes over the internal geometry generated by the $\nc$ D3-branes. In other words, we consider a  distribution of D7-branes that share all the D3-brane directions and are homogeneously smeared along the internal directions. On the gauge theory side the smearing means that we are coupling the original super-conformal theory to quarks with all possible R-symmetry quantum numbers compatible with $\mathcal{N}=1$ supersymmetry. The technical advantages of the smearing procedure are that it reduces the problem to a codimension-one problem on the gravity side\cite{Bigazzi:2005md}, and that it enlarges the regime of validity of the supergravity solution, as we will see in Sec.~\ref{sec.susysolution}. 

The supersymmetric  solution describing the ground state of the D3-D7 system was constructed analytically in \cite{Benini:2006hh} and we will review it in 
Sec.~\ref{sec.susysolution}. As expected, the solution possesses a singularity at which  the dilaton diverges located at a finite distance along the holographic coordinate. This is interpreted as the gravity dual of the Landau pole singularity in the coupling constant of the gauge theory that occurs at a finite energy scale. The key result from Sec.~\ref{sec.susysolution} is that, near the Landau pole singularity,  the effective five-dimensional metric obtained by reduction along the five internal directions takes the form of a hyper-scaling violating (HV) metric with HV exponent $\theta > p$, where $p=3$ is the number of spatial directions of the gauge theory. All the qualitative manifestations of the Landau pole that we study in subsequent sections follow from this fact. 

In Sec.~\ref{LPphysics} we study these manifestations in the ground state solution of  \cite{Benini:2006hh}. We show that the presence of the Landau pole manifests itself, for example, in a maximum in the possible density of degrees of freedom per unit volume in the gauge theory. In this section we also determine the non-trivial map between the holographic coordinate and the energy scale in the gauge theory, which we use to compute the $\beta$-function. In Sec.~\ref{qqsec} we show that the existence of a UV cut-off in the gauge theory is also visible in the behavior of the quark-antiquark potential and the entanglement entropy (EE). 

In Sec.~\ref{finiteT} we turn to the numerical construction of solutions with non-zero temperature.  In Ref.~\cite{Bigazzi:2009bk} these solutions were constructed perturbatively in $T/\lp$, and are therefore valid only at low temperatures. We construct solutions valid at any temperature and use them in Sec.~\ref{app.neutralthermo} to study the thermodynamics of the system. A crucial ingredient in this analysis is the implementation of the holographic renormalization procedure, which allows us to define a finite free energy, a finite boundary stress tensor, etc.\footnote{
%DMDM
Holographic renormalization is best understood for conformal theories. Studies of non-conformal cases include non-conformal branes \cite{Kanitscheider:2008kd,Wiseman:2008qa}, HV Lifshitz theories \cite{Papadimitriou:2011qb,Chemissany:2014xsa} and cascading theories \cite{Aharony:2005zr,Bertolini:2015hua}.} Following \cite{Kanitscheider:2009as,Gouteraux:2011ce} we show in Appendix \ref{continuation} that the near-Landau pole HV metric can be related to an AdS metric through analytic continuation in the number of dimensions. As a result the holographic renormalization of AdS can be continued back to define the procedure in full generality for the HV metric. The main result of this section is that the solutions become locally thermodynamically unstable at a temperature $T \lesssim \lp$ because  they develop a negative specific heat.  Interestingly though, before this happens, i.e.~in the stable phase, the speed of sound exceeds the conformal value, $c_s^2 > 1/3$. 

We close in Sec.~\ref{disc} with a discussion of our results and a possibility suggested by recent work in Quantum Electrodynamics.

%%%%%%%%%%%%%%%
%%%%%%%%%%%%%%%%%%%%%%%%%%%%%%%%
\section{Preliminaries}\label{sec.setup}
%%%%%%%%%%%%%%%%%%%%%%%%%%%%%%%%
%%%%%%%%%%%%%%%

Four-dimensional, $\mathcal{N}=4$ SYM theory with $\nc$ colors is holographically dual to supergravity solutions sourced by $\nc$ D3-branes. In the supergravity description this is encoded in a flux of the self-dual RR five-form through an appropriate five-dimensional compact manifold $\cM_5$,
\be
F_5 = \Qc (1+*) \omega_5 \,,
\ee
where $\omega_5$ is the volume form of $\cM_5$, whose total dimensionless volume we denote $V_5$.  Quantization requires that the D3-brane charge is related to the number of colors through
\be
\label{qqcc}
\Qc = \frac{(2\pi\ls)^4}{2\pi V_5} \, \nc \ .
\ee
Note that, unlike in Refs.~\cite{Faedo:2014ana,Faedo:2015ula,Faedo:2015urf,Faedo:2016jbd}, we work with an RR charge quantized in units of $\nc$ instead of $g_s \nc$ (for a comparison between both normalizations, see e.g.~Sec.~4.1 of \cite{Mateos:2011tv}). The latter choice is convenient in situations in which there is natural factorization of the dilaton $\phi$ of the form $e^\phi=g_s e^{\tilde \phi}$. This is the case, for example, if the gauge theory is conformal, since this means that the dilaton is constant and one can simply normalize it so that $\tilde \phi=0$ everywhere. If the gauge theory is not conformal but approaches a fixed point in the infrared (IR) or in the UV then it is natural to normalize the dilaton so that $\tilde \phi=0$ at the corresponding fixed point. In contrast, in the solutions that we will consider the dilaton  will run from zero to infinity and there will be no natural factorization  into a constant piece and a running piece. We will therefore work with the full dilaton, which is related to the running YM and 't Hooft couplings through 
\be
\gym^2 = 2\pi e^\phi \sac \lambda =  \gym^2 \nc \,.
\ee

In the simplest cases  the metric supported by the $F_5$  flux is AdS$_5\times\cM_5$, with $\cM_5$ a Sasaki-Einstein (SE) manifold. The radii of these two spaces is related to the D3-brane charge through
\be
\label{correct}
\Qc = 4 L^4 \,. 
\ee
If \mbox{$\cM_5=S^5$} then the gauge theory is $\cN=4$ SYM; otherwise it is a non-maximally supersymmetric theory. For example, if $\cM_5=T^{1,1}$ then the gauge theory  is the Klebanov-Witten quiver. In the general case the rank of the gauge group and $L$ are related through 
\be
\label{kkappa2}
\frac{L^3}{\kappa_5^2} = \frac{\pi \nc^2}{4 V_5}  \,,
\ee
where $\kappa_5^2$ is the five-dimensional effective gravitational coupling (see \eqq{eq.neutral5daction} below).

In order to add flavor to any of the theories above it is convenient to view the SE manifold as a  U(1) fibration over a four-dimensional  K\"ahler--Einstein (KE) base. This geometric construction is naturally equipped with an SU(2) structure characterized by a real  one-form, $\etakah$, and a real two-form, $\Jkah$, which is  the K\"ahler form of the  KE manifold. These satisfy the relations
\be\label{eq.SU2structure}
\d \etakah = 2 \Jkah \ , \qquad  \frac{1}{2}\Jkah\wedge \Jkah \wedge \etakah = \omega_5 \ .
\ee
They also close on each other under Hodge dualization on the SE manifold 
\be\label{eq.SU2closing}
*_5 \etakah = \frac{1}{2} \Jkah \wedge \Jkah \ , \qquad *_5 \Jkah = \Jkah \wedge \etakah \ .
\ee
The addition of flavor on the gauge theory corresponds on the gravity side to the addition  of $\nf$ D7-branes, as explained in Sec.~\ref{intro}. Their presence is encoded in a flux of the  RR one-form 
\be
F_1 = \Qf\, \etakah \ ,
\ee
where the D7-brane charge $\Qf$ is related to the number of D7-branes through
\be
\Qf = \frac{V_3 }{8\pi V_5}  \, \nf \,,
\label{qqff}
\ee
with $V_3 = \int \Jkah \wedge \etakah$ the dimensionless volume of the three-dimensional submanifold $\cM_3\subset \cM_5$ wrapped by any of the D7-branes. Note that the smeared D7-branes lead to a violation of the Bianchi identity for $F_1$, namely to the fact that $\d F_1\neq 0$, as expected from a continuous distribution of objects charged under $F_1$ \cite{Benini:2006hh}.

The ten-dimensional string- and Einstein-frame  metrics of the flavored solution take the form 
\be
\label{eq.10dmetric}
\d s^2_\mt{st} \,=\,  e^{\phi/2} \d s^2_\mt{E} \,= \, e^{\phi/2} 
\left[ \sfh^{-1/2} \left( -\sfb\, \d t^2 + \d  x_3^2 \right) + \sfh^{1/2} e^{2 \sff} \sqrt{\Qc} \left( \frac{\d \varrho^2}{\sfb\, \sfc}  + e^{2\sfg-2\sff} \d s_\mt{KE}^2 + \etakah^2 \right) \right] \,.
\ee
The undetermined functions in this expression depend only on the radial coordinate $\varrho$. Note that this coordinate is dimensionless, consistent with  the fact that $\Qc$ has dimensions of (length)$^4$. We have included a blackening factor $\sfb$ since below we will consider solutions with non-zero temperature, as well as an additional function $\sfc$ since we want to fix the radial gauge dynamically. We have also allowed for a relative squashing between the  KE base and the fiber, proportional to $ e^{2\sfg-2\sff}$, which will only be absent  in the case without flavor. 

It is convenient to reduce the ansatz above along the internal $\cM_5$ compact manifold. The result is the five-dimensional effective action\footnote{In the case without flavor this reduction was done in \cite{Klebanov:2000nc,Benvenuti:2005qb}.} \cite{Benini:2006hh,Cotrone:2012um} 
\be\label{eq.neutral5daction}
S_{5} = \frac{1}{2 \kappa_5^2} \int (R-V)*1 - \frac{4}{5} \d \chi \wedge * \d \chi - \frac{8}{15} \d \psi \wedge * \d \psi - \frac{1}{2} \d \phi \wedge * \d \phi \,.
\ee 
The scalars $\chi, \psi$ in this action are related to the functions  $\sff, \sfg$ parametrizing the internal manifold as
\be\label{eq.newscalars}
e^\psi = \sfh^{5/4} e^{4\sfg+\sff}  \  , \qquad e^\chi = e^{\sff-\sfg} \,,
\ee
and the five-dimensional metric is related to the ten-dimensional  one through 
\be
\d s^2 = e^{\phi/2} \left[ e^{-\frac{2}{3}\psi}\d s_5^2 + \sqrt{\Qc}\,  e^{-\frac{2}{5}(\chi-\psi)} \d s_\mt{KE}^2 + \sqrt{\Qc} \, e^{\frac{2}{5}(4\chi+\psi)} \etakah^2 \right] \,,
\ee
namely
\be
\label{eq.5dmetricdef}
\d s_5^2 = e^{\frac{2}{3}\psi} \left[ \sfh^{-1/2} \left( -\sfb\, \d t^2 + \d  x_3^2 \right) + \sfh^{1/2} e^{2 \sff} \sqrt{\Qc}  \, \frac{\d \varrho^2}{\sfb\, \sfc} \right] \,.
\ee
Although in most cases it will be clear from the context, to avoid any possible confusion we will denote the ten-dimensional metric as $G$ and the five-dimensional metric as $g$. If necessary we will also write $G^\mt{st}$ and 
$G^\mt{E}$ to distinguish between the string- and the Einstein-frame metrics.

The potential in the five-dimensional action is derived from the superpotential 
\be\label{eq.superpotential}
W = \frac{e^{-\frac{8}{15}\psi}}{2\cdot 2^{1/3}L} \left[ e^{-\frac{4}{5}\psi} 
- e^{- \frac{4}{5}\chi} \left( 6 + 4 \, e^{2\chi} - \Qf \, e^\phi \right) \right] 
\ee
via the standard relation
\be
\label{potpot}
V = \frac{1}{2} \left[ \frac{5}{8} \left( \frac{\partial W}{\partial \chi} \right)^2 + \frac{15}{16} \left( \frac{\partial W}{\partial \psi} \right)^2 +  \left( \frac{\partial W}{\partial \phi} \right)^2 \right] - \frac{W^2}{3} \,.
\ee
This potential encodes the effects of the two RR kinetic terms in the IIB supergravity action as well as the smeared Dirac-Born-Infeld term describing the presence of the flavor D7-branes. 

The equations of motion for the entire system can be easily obtained as the Euler-Lagrange equations associated to the action \eqref{eq.neutral5daction}, and can be found in the references \cite{Benini:2006hh,Nunez:2010sf,Bigazzi:2009bk,Benini:2007kg,Cotrone:2012um}.

%%%%%%%%%%%%%%%
%%%%%%%%%%%%%%%%%%%%%%%%%%%%%%%%
\section{Supersymmetric solution}\label{sec.susysolution}
%%%%%%%%%%%%%%%%%%%%%%%%%%%%%%%%
%%%%%%%%%%%%%%%

In this section we will use the ten-dimensional  parameterization of the metric given in \eqref{eq.10dmetric} with $\sfb=\sfc=1$ and $\sfg$ traded in favor of $\chi$ as in \eqref{eq.newscalars}. The D3-D7 system with smeared D7-branes preserves $\cN=1$ supersymmetry and the corresponding BPS equations are  \cite{Benini:2006hh}
\be\label{eq.BPSeqs}\bal
\partial_\varrho \phi & = -\Qf\, e^{\phi} \ , 
& \partial_\varrho \chi  =  -3 \left( 1- e^{2\chi} \right) + \frac{1}{2} \Qf \, e^\phi \ , \\
\partial_\varrho \sff & =  -3 + 2 e^{2\chi} + \frac{1}{2} \Qf \, e^\phi \ , 
\qquad & \partial_\varrho \sfh  =  e^{4(\chi-\sff)} \ .
\eal\ee
When $\Qf=0$ this has the well known AdS$_5\times\cM_5$ solution, with $\partial_\varrho \phi = 0$ and $\chi = 0$. In the flavored case the solution reads
\be\label{eq.susysolution}
\bal
e^\phi & = \frac{e^{\phi_0}}{1+\Qf\, e^{\phi_0}(\varrho-\varrho_0)} \ , \\[4mm]
e^\chi & = \frac{ 6^{1/2} \sqrt{1+\Qf\,  e^{\phi_0} ( \varrho - \varrho_0) }}{ \sqrt{6 + \Qf\, e^{\phi_0} \left[ 1+6(\varrho - \varrho_0)  + c_1\, e^{6/(\Qf\, e^{\phi_0})}\,  e^{ 6(\varrho -\varrho_0 )}  \right] } }  \ , 
\\[5mm]
e^\sff & = \frac{c_2\, e^{-1/ ( \Qf \, e^{\phi_0} ) }}{\left( \Qf \, e^{\phi_0} \right)^{1/6}} \frac{6^{1/2} e^{-(\varrho-\varrho_0)} 
\sqrt{1+ \Qf \,e^{\phi_0} (\varrho-\varrho_0)} }
{ \Big[ 6 + \Qf\, e^{\phi_0} \left[ 1+6(\varrho - \varrho_0) + c_1\, e^{6/(\Qf\, e^{\phi_0})}\,  e^{ 6(\varrho -\varrho_0 )}  \right] \Big]^{1/3} }   \ , \\[4mm]
\sfh & =c_3 +  \int^\varrho 
\, e^{4 \chi(\hat \varrho)} \, e^{-4 \sff(\hat \varrho)} \, \d \hat \varrho  \ ,
\eal
\ee
where $\phi_0 \equiv \phi(\varrho_0)$ is the value of the dilaton at  an arbitrary reference point $\varrho_0$, and $c_{1,2,3}$ are constants of integration.  In order for the dilaton to be real the radial coordinate must satisfy 
\be
\varrho_\mt{LP}  <  \varrho
\ee
where 
\be\label{eq.varrhoLP}
 \varrho_\mt{LP}\equiv \varrho_0 - \frac{1}{\Qf \, e^{\phi_0}}  \,.
\ee
At precisely $\varrho_\mt{LP}$ the dilaton diverges, in agreement with  the presence of a Landau pole at which  the coupling constant diverges. Using \eq{eq.varrhoLP} one may rewrite the solution in terms of $\rholp$ alone as 
\be\label{eq.susysolution2}
\bal
e^\phi  &= \frac{1}{\Qf\, (\varrho-\varrho_\mt{LP})} \ , \\[4mm]
e^\chi & = 6^{1/2} \sqrt{\frac{\varrho - \rholp}{1+6(\varrho - \rholp)+ c_1 \, e^{6(\varrho-\rholp)} }} \,,
\\[5mm]
e^\sff & = c_2 \, 6^{1/2} \,e^{-(\varrho-\rholp)}
\frac{\sqrt{ \varrho-\rholp}}{\left[ 1+6 (\varrho - \rholp) + c_1 \, e^{6(\varrho-\rholp)} \right]^{1/3}}  \,, \\[4mm]
\eal
\ee
with $\sfh$ as in \eq{eq.susysolution}. We see that by choosing the Landau pole as the reference point the solution takes a particularly simple form. By an additional shift of the radial coordinate we will set $\rholp=0$ without loss of generality. 

Let us now discuss the role of the integration constants. As noted in \cite{Benini:2006hh}, if the integration constant $c_1$  in \eqref{eq.susysolution} is non-zero then  the proper radial distance from any point to the IR of the theory is finite, whereas if $c_1=0$ there is an infinite throat. We will therefore set $c_1=0$ hereafter, since otherwise there would be an unphysical IR cut-off in the gauge theory. This means, in particular, that the range of the radial coordinate is $0  <  \varrho < \infty$. We will refer to the region  $\varrho\to\infty$ as the IR part of the geometry. In this region the dilaton approaches zero, as expected from the fact that the dual gauge theory is IR-free \cite{Aharony:1998xz}, and the curvature invariants diverge, as we discuss below. This divergence is absent if the quark mass is non-zero \cite{Nunez:2010sf}, as we discuss in Sec.~\ref{disc}. 

The constant $c_2$ can be set to any convenient value through the following rescalings of $c_3$ and of the Minkowski coordinates
\be\label{eq.minkowskirenorm}
x^\mu \to x^\mu\,  c_2^{-1} \ , \qquad c_3 \to c_3\, c_2^{-4} \ .
\ee
Two particularly convenient choices are $c_2=1$ and $c_2=\Qf^{1/6}$. With the former choice the  solution takes a particularly simple form in the UV, as we will see below, whereas with the latter  the flavorless limit $\Qf \to 0$ in \eqref{eq.susysolution} becomes manifestly smooth. 

Setting $c_1=0$ and redefining the last integration constant $c_3$ appropriately, the warp factor takes the form
\be\label{eq.hsupersymmetric}
\sfh =c_3 + \left( \frac{-1}{2e} \right)^{2/3} 
\frac{6^{-2/3}}{c_2^4}
\left(  \Gamma \left[\frac{1}{3}, -\frac{2}{3} - 4 \varrho \right]  -\Gamma \left[ \frac{1}{3}, -\frac{2}{3} \right]  \right)  \ ,
\ee
where $\Gamma[s,x]$ is the incomplete Gamma function. Since the arguments of the incomplete Gamma functions are always negative and
\be
\text{Im}\,  \left\{ (-1)^{2/3} \, \Gamma\left[\frac{1}{3},-x \right] \right\}
 = \frac{\sqrt{3}}{2} \, \Gamma\left( \frac{1}{3} \right) \qquad \text{ if }x>0 \,,
\ee
it follows that  $\sfh$ is real if $c_3$ is real.  For the flavored case $\sfh$ vanishes linearly as $\varrho \to 0$  if $c_3=0$ and it approaches a constant  if $c_3\neq 0$. Following \cite{Benini:2006hh} hereafter we will set $c_3=0$. This is analogous to ``dropping the 1'' in the usual warp factor of D3-branes. It would be interesting to extend our analysis to the case in which $c_3\neq 0$, which would correspond to studying smeared D7-branes in the asymptotically flat  background sourced by D3-branes.

\subsection{IR and UV asymptotics}

Having fixed the solution, let us now examine its IR and UV limits. In the IR region we have $\varrho \gg 1$ and the  metric components  asymptote to
\be\label{eq.leadingIRsusy}
\mbox{IR:} \qquad
e^\phi  \simeq \frac{1}{\Qf\, \varrho} \ ,\qquad
e^\chi  \simeq 1 \ , \qquad 
e^\sff  \simeq c_2\,  6^{1/6}\, \varrho^{1/6}\, e^{-\varrho} \ , \qquad
\sfh  \simeq \frac{e^{4\varrho}\, \varrho^{-2/3}}{4\cdot 6^{2/3} c_2^4}\ .
\ee
The asymptotic form of the solution is most transparently understood in terms of a new radial coordinate $z$ defined through 
\be
\label{IRcoordinate}
\varrho=\log \frac{z}{L}\,.
\ee
In this  coordinate the IR region is at $z\to \infty$ and  the asymptotic form of the five-dimensional metric \eq{eq.5dmetricdef} is
\be
\label{IRmetric}
\d s^2_5 \simeq \frac{1}{2^{8/3}} \left[  4\cdot 6^{1/3} c_2^2 \left( \frac{L}{z} \right)^2 \left( \log \frac{z}{L}\right)^{1/3} \eta_{\mu\nu} \d x^\mu \d x^\nu + L^2 \frac{\d z^2}{z^2} \right] \ .
\ee
Were it not for the presence of the logarithmic term this would be simply AdS$_5$ spacetime. This term is a deformation from this spacetime due to the presence of flavor in the system.  We will  refer to this metric as ``logarithmically-corrected AdS'' or simply ``log-AdS''. As we will show below, this geometry behaves essentially as AdS for probes that do not explore exponentially large distances along the $z$ coordinate. 

The metric \eq{IRmetric}  is accompanied by the three five-dimensional scalars whose asymptotic behavior is 
\be
e^\phi \simeq  \frac{1}{\Qf \, \log ( z/L)} \ , \qquad 
e^\psi \simeq   \frac{1}{4\sqrt{2}} \ , \qquad 
e^\chi \simeq  1 \,.
\ee
We see that asymptotically the squashing of the internal manifold disappears since $e^\chi \simeq1$, and  the size of the compact part of the geometry approaches a  finite value. The dilaton vanishes logarithmically slowly towards the IR. As we will elaborate upon below, this  is intimately related to the logarithmic running of the YM coupling of the dual gauge theory at energies much below the Landau pole. 

In the UV we have $\varrho\to 0$ and the solution becomes
\be\label{eq.leadingUVsusy}
\bal
\mbox{UV:} \qquad e^\phi & = \frac{1}{\Qf\, \varrho} \ , \\[3mm]
e^\chi & \simeq \sqrt{6 \varrho}\, \Big[ 1 - 3 \varrho + \cdots \Big] \ , \\[2mm]
e^\sff & \simeq  c_2\,\sqrt{6 \varrho}\, 
\Big[ 1 - 3 \varrho  + \cdots \Big] \ , 
\\[2mm]
\sfh & \simeq \frac{\varrho }{c_2^4} \,  \Big[ 1 +4 \varrho^2 + \cdots \Big]\ .
\eal\ee
The asymptotic form of the solution is most transparently understood in terms of a new radial coordinate $u$ defined through
\be
\label{UVcoordinate}
\varrho = \left( \frac{L}{u} \right)^{1/2} \,.
\ee
In this  coordinate the Landau pole is at $u\to \infty$ and  the asymptotic form of the five-dimensional metric in this region is
\be\label{eq.LandaupoleasHV}
\d s^2 \simeq  3\cdot 6^{1/3} \, \left(\frac{u}{L}\right)^{-\frac{2\theta}{p}} \left(  
\frac{c_2^2}{3} \,  \frac{u^2}{L^2} \, \eta_{\mu\nu}\d x^\mu \d x^\nu +  \frac{L^2}{u^2} \d u^2 \right) \,,
\ee
where $p=3$ is the number of spatial dimensions and $\theta=7/2$. 
This geometry is accompanied by the three five-dimensional  scalars which, up to subleading corrections of order $u^{-1/2}$, take the form
\be
e^\phi \simeq  \frac{1}{\Qf} \left( \frac{u}{L} \right)^{1/2} \ , \qquad 
e^\psi \simeq   6^{1/2} \left( \frac{u}{L} \right)^{-7/8} \ , \qquad 
e^\chi \simeq  6^{1/2} \left( \frac{u}{L} \right)^{-1/4} \,.
\ee
The metric \eq{eq.LandaupoleasHV} is a HV metric with HV parameter $\theta$  \cite{Ogawa:2011bz,Huijse:2011ef,Dong:2012se}. As we will see, all the qualitative manifestations of the Landau pole that we study in subsequent sections follow from this fact, which to our knowledge had not been noted before. The overall  factor $3  \cdot  6^{1/3}$ and the $c_2^2/3$ factor inside the brackets could be eliminated by appropriate rescaling of the length scale $L$ and of the Minkowski coordinates. However, we will not do so and  work with the asymptotic normalization of gauge theory coordinates given in  \eq{eq.LandaupoleasHV}. Changing this normalization would, for example, rescale our definition of temperature below. 

%JTJT
Models with $\theta>p$ have appeared previously in the literature (see 
e.g.~\cite{Amoretti:2016cad} for a recent example) but in circumstances in which Lorentz invariance is broken and in which they control the IR instead of the UV physics of the theory. 

\subsection{Regime of validity}

\begin{figure}[t]
\begin{center}
\includegraphics[width=0.45\textwidth]{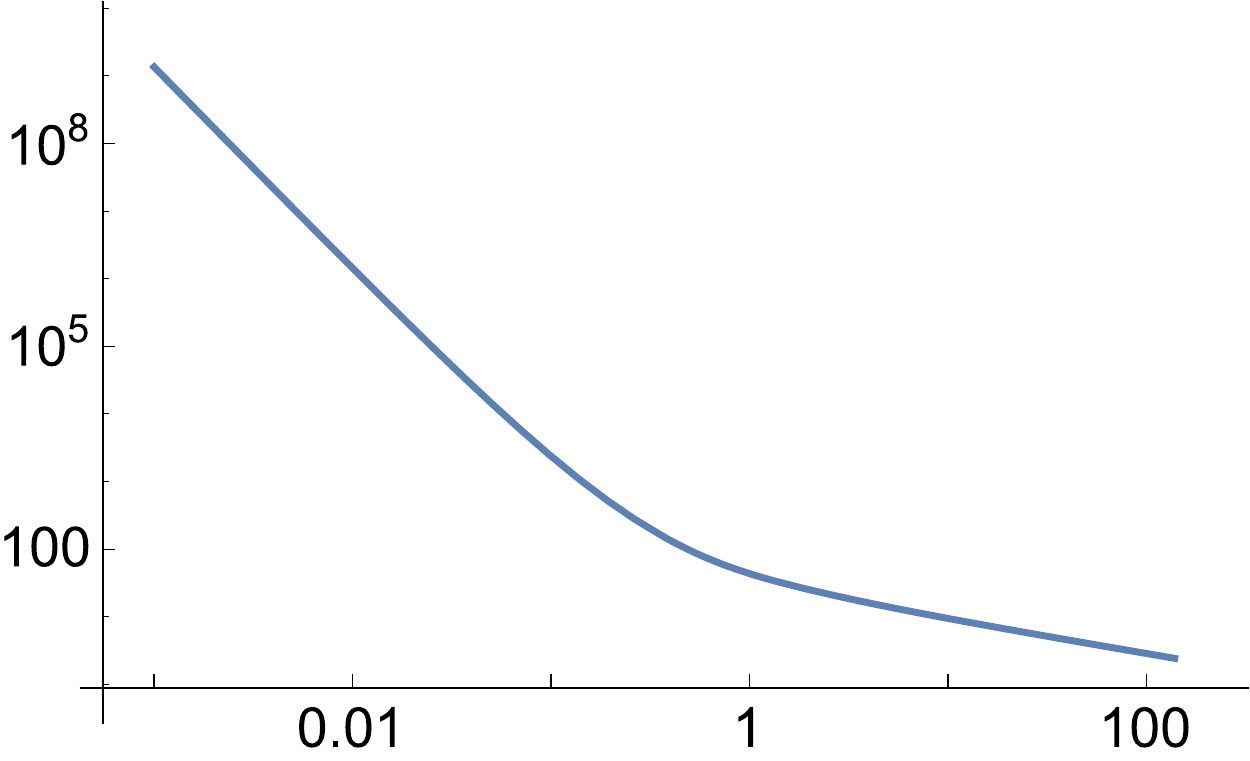} \qquad
\includegraphics[width=0.45\textwidth]{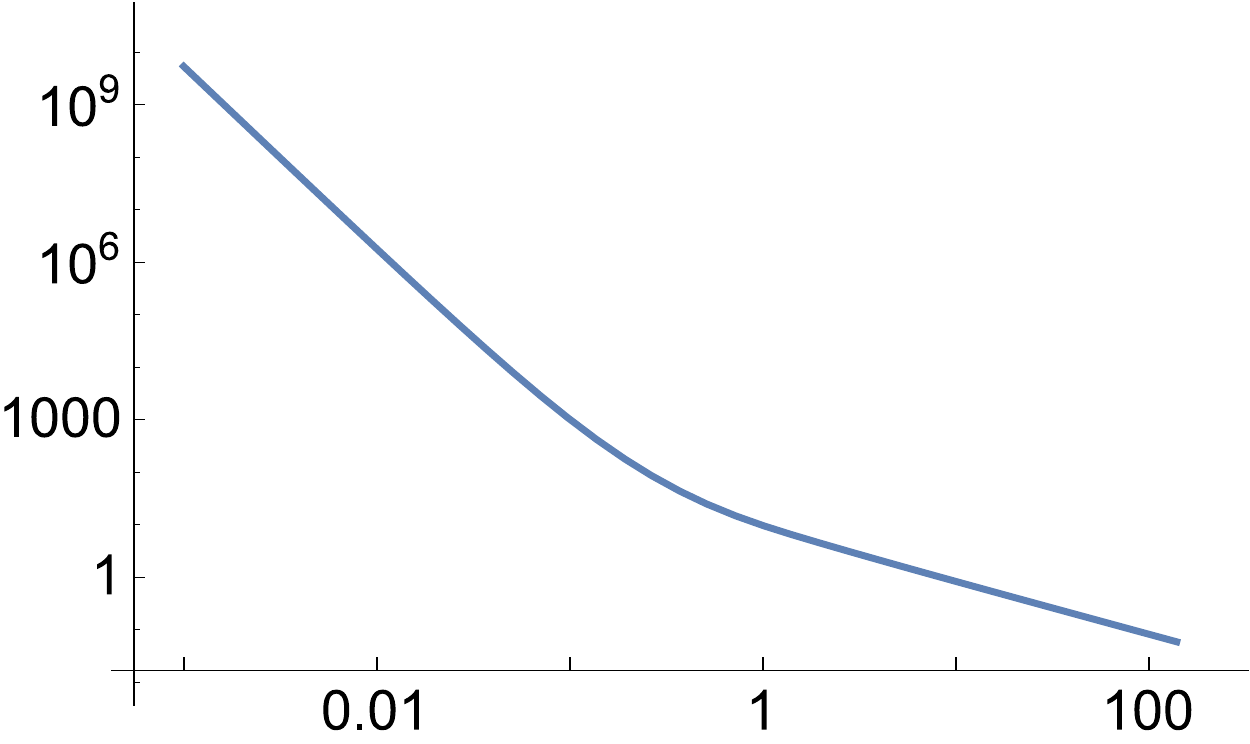}  
\put(-200,125){ $\sqrt{\Qc}\, \mbox{R}_\mt{Ein}$}
\put(5,10){ $\varrho$}
\put(-425,125){ $\sqrt{\Qc/\Qf}\, \mbox{R}_\mt{st}$}
\put(-220,10){ $\varrho$}\\[8mm]
\includegraphics[width=0.45\textwidth]{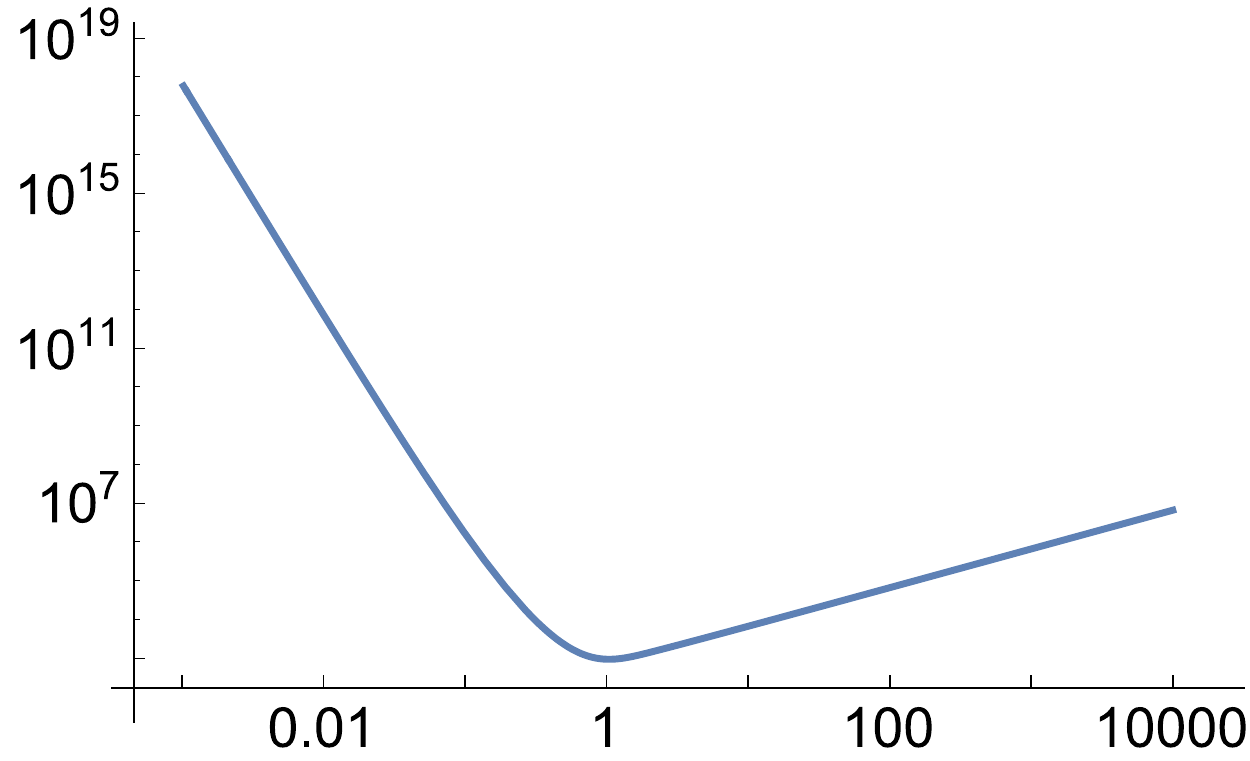} \qquad
\includegraphics[width=0.45\textwidth]{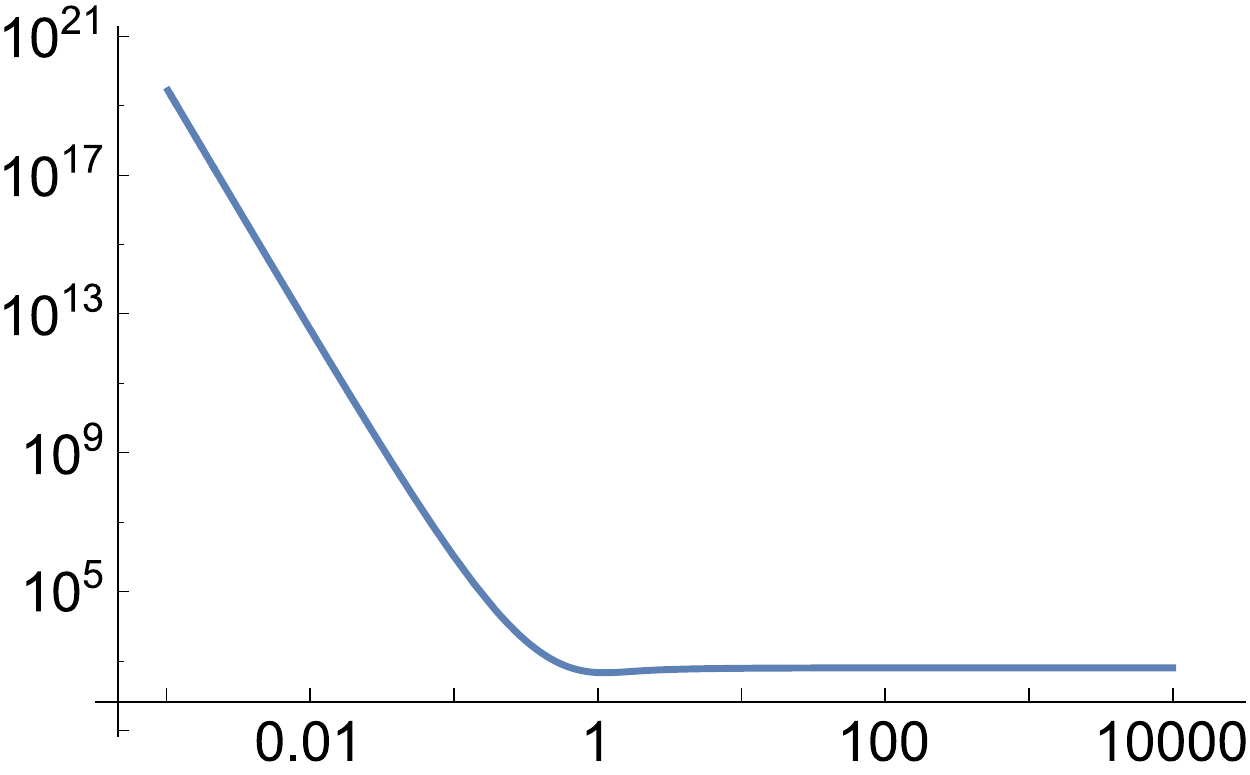}  
\put(-200,130){ $\Qc\, \mbox{Ric}_\mt{Ein}^2$}
\put(5,10){ $\varrho$}
\put(-425,130){ $\Qc/\Qf\, \mbox{Ric}_\mt{st}^2$}
\put(-220,10){ $\varrho$}
\caption{Curvatures of the string-frame (left) and Einstein-frame (right) ten-dimensional metrics.}
\label{curv}
\end{center}
\end{figure}
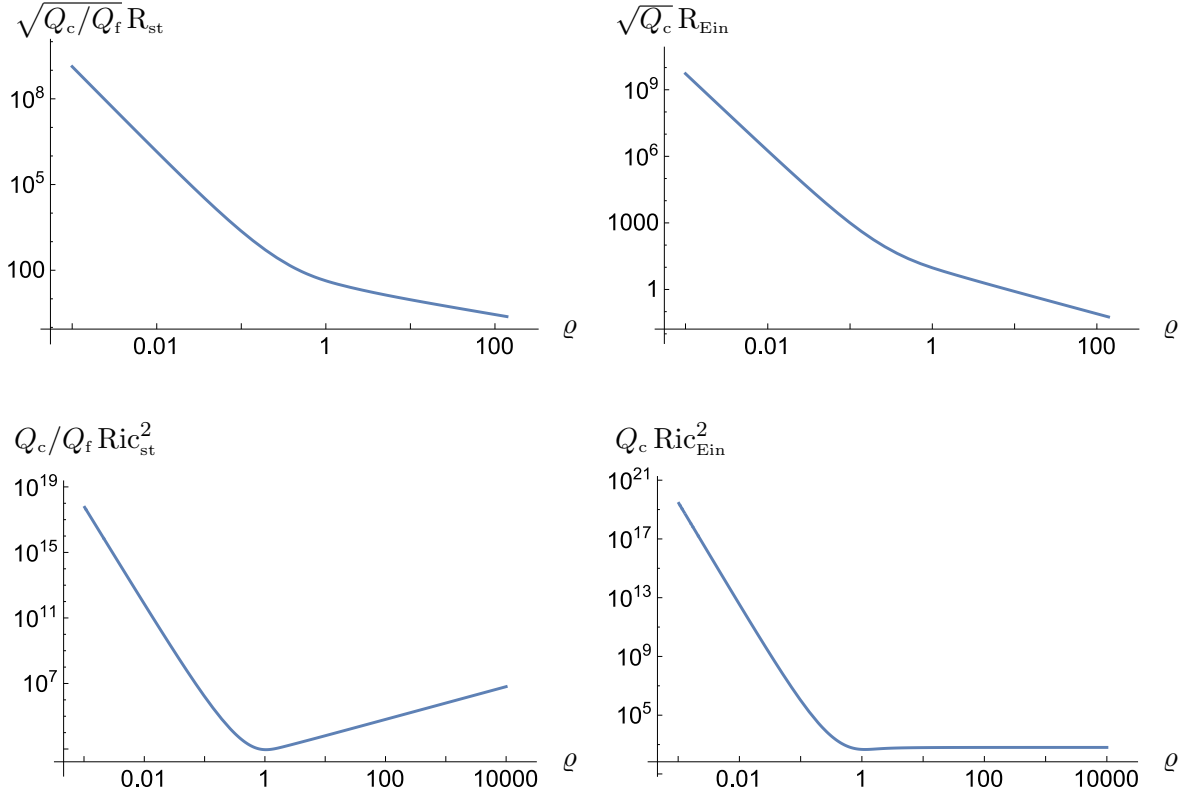

We are now ready to determine the regime of validity of the solution. Since we are only interested in parametric dependences, in this analysis we will ignore all purely numerical factors. We must require that both supergravity and the smeared DBI action for the D7-branes be valid. We begin with supergravity. In this case the first condition that we must impose is 
\be
e^\phi \ll 1 \,.
\label{firstcondition}
\ee
If this  is not satisfied then string loop corrections become important and degrees of freedom not included in supergravity, such as D-branes, become light. The other two conditions that we must require are that the curvature of the string-frame metric be small in string units, and that the curvature of the Einstein-frame metric be small in Planck units.\footnote{Note that in our conventions $\ell_p^4 \sim \ell_s^4$ because we are not factoring the dilaton into a constant times a position-dependent part.} The result can be understood simply by considering the two asymptotic forms of the solution in the IR and in the UV. The reason is that, as shown in 
Fig.~\ref{curv}, the curvatures exhibit two simple behaviors separated by a rapid crossover  around $\varrho \sim 1$. 
We will now see that these  behaviors are controlled by the IR and the UV asymptotic solutions. Since the Ricci scalar, $\mbox{R}=G^{mn} R_{mn}$, and the square of the Ricci tensor, $\mbox{Ric}^2=R^{mn} R_{mn}$, behave differently, we need to consider both. We have checked that the square of the Riemann tensor, $R^{mnpq} R_{mnpq}$, does not give new independent conditions;  presumably the same is true for other curvature invariants. Through explicit calculation we find that the conditions that the string-frame curvatures be small in string units take the form 
\bea
\label{cond1}
\mbox{IR:} \qquad \ell_s^4 \,  \mbox{R}_\mt{st}^2 &\sim& \frac{\nf}{\nc} \, e^\phi 
\,\,\, \ll \,\,\, 1\,,
\\[2mm]
\label{cond2}
\ell_s^4 \, \mbox{Ric}_\mt{st}^2 &\sim& \frac{1}{\nc \, e^\phi}  
\,\,\, \ll \,\,\, 1 \,, 
\\[2mm]
\label{cond3}
\mbox{UV:} \qquad \ell_s^4 \,  \mbox{R}_\mt{st}^2 &\sim& 
\ell_s^4 \, \mbox{Ric}_\mt{st}^2 \sim 
\frac{\nf^7}{\nc} \, e^{6\phi} \,\,\, \ll \,\,\, 1 \,.
\eea
The curvatures of the Einstein-frame metric in Planck units turn out to be proportional to the string-frame curvatures in string units,  both in the IR and in the UV:  
\be
\ell_p^4 \, \mbox{R}_\mt{Ein}^2 \sim e^{\phi} \, \ell_s^4 \,  \mbox{R}_\mt{st}^2  \sac
\ell_p^4 \, \mbox{Ric}_\mt{Ein}^2 \sim e^{\phi} \, \ell_s^4 \,  \mbox{Ric}_\mt{st}^2  \,.
\ee
It follows that the condition \eq{firstcondition} and the requirement that the string-frame curvatures be small in string units imply that the Einstein-frame curvatures are small in Planck units. Therefore we will ignore the Einstein-frame curvatures in the following. 

Using the dilaton profile 
\be
\label{profile}
e^\phi \sim \frac{1}{\nf  \, \varrho } 
\ee
in Eqs.~\eq{cond1}-\eq{cond3} it is easy to check that the $\varrho$-dependence of the curvatures in the IR and in the UV regimes matches the $\rho \gg 1$ and the $\rho \ll 1$ behaviors shown in Fig.~\ref{curv}, respectively. Moreover, using the dilaton profile and the conditions \eq{firstcondition}-\eq{cond3} one can determine the range of the $\varrho$ coordinate where the solution is reliably described by supergravity. The point $\varrho\sim 1$  is particularly relevant since, as we will see below, several interesting features of the solution begin  to manifest themselves around this point. In terms of the $\varrho$ coordinate the dilaton condition implies
\be
\label{stringent2}
\frac{1}{\nf} \ll  \varrho \,.
\ee
Since our construction based on smeared D7-branes requires $\nf \gg 1$, we see that this condition is satisfied at $\varrho\sim 1$. This means that the region where \eq{firstcondition} holds includes parts of both the IR and the UV regions. Therefore we must make sure that the curvatures are small in both regions. In the IR, the condition \eq{cond2} translates into
\be
\varrho \ll \frac{\nc}{\nf} \,,
\ee
whereas the UV condition \eq{cond3} implies
\be
\label{stringent1}
\left( \frac{\nf}{\nc} \right)^{1/6} \ll \varrho \,.
\ee
We see that if $\nf / \nc \sim 1$ then there is essentially no region where the  supergravity solution is valid. We will therefore assume that 
\be
\frac{\nf}{\nc} \ll 1 \,.
\ee
Since $\nf \gg 1$ this automatically implies $\nc \gg 1$. Under these circumstances the IR condition \eq{cond1} is automatically satisfied. 
We will see below that \eq{stringent1} may be more or less stringent than \eq{stringent2}. 

To summarize, supergravity is a valid description over a significant region provided we have
\be 
\label{stringent}
1 \ll \nf \ll \nc \,.
\ee
Under these circumstances, in terms of the supergravity coordinate this region is given by 
\be
\label{low}
\varrho_\mt{UV}  \ll \varrho \ll \varrho_\mt{IR} \sac 
\varrho_\mt{IR} = \frac{\nc}{\nf} \sac 
\varrho_\mt{UV} = \max \left\{ \frac{1}{\nf} , \left( \frac{\nf}{\nc} \right)^{1/6} \right\} \,,
\ee
whereas in terms of the gauge theory coupling we have
\be
\gym^2 \ll 1 \sac 1 \ll \lambda \ll \min \left\{ 
\nc , \left( \frac{\nc}{\nf} \right)^{7/6} \right\} \,,
\ee
where $\lambda = \gym^2 \nc$.  

We now turn to  the constraints imposed by the requirement that the Abelian DBI action for the D7-branes be valid \cite{Bigazzi:2008zt,HoyosBadajoz:2008fw}. The first requirement concerns the characteristic distance between nearby D7-branes, and it consists of two complementary conditions. On the one hand, this distance must be small in macroscopic terms in order for the distribution to be treated as continuous. This simply implies that  $\nf \gg 1$. On the other hand, this distance must be large in string units, since otherwise strings stretching between nearby D7-branes would become light and 
the non-Abelian nature of the DBI action would become important. Since all the D7-branes wrap the $\eta$-fiber in the internal geometry of the metric \eq{eq.10dmetric} we must consider their separation in the KE base. The characteristic size of this manifold is 
\be
\ell = \sqrt{G^\mt{st}_\mt{KE}} = \Qc^{1/4} e^{\phi/4} h^{1/4} e^{f} e^{-\chi} \,.
\ee
Since inside the four-dimensional KE base the branes are co-dimension two objects, one may effectively think of them as points in a two-dimensional space of volume $\sim \ell^2$. The volume available to each of the branes is therefore $\ell^2/\nf$. As a consequence, the typical inter-brane distance is 
$\sqrt{\ell^2/\nf}$. The requirement that this distance is large in string units is thus
\be
\label{inter}
\frac{\ell^4}{\nf^2\, \ell_s^4} \gg 1\,.
\ee
Using \eq{eq.leadingIRsusy} and \eq{eq.leadingUVsusy} we find the leading asymptotic behaviors 
\bea
\label{using}
\mbox{IR:} \qquad && \ell^4\sim \frac{\nc \, \ell_s^4}{\nf \, \vr} \,, \\[2mm]
\mbox{UV:} \qquad && \ell^4\sim \frac{\nc \, \ell_s^4}{\nf} \,.
\eea
Substituting into \eq{inter} yields the two conditions
\bea
\mbox{IR:} \qquad && \frac{\nc}{\nf^3}\,  \frac{1}{\vr} \gg 1  \,, \\[2mm]
\mbox{UV:} \qquad && \frac{\nc}{\nf^3} \gg 1 \,.
\eea
We see that the IR condition immediately implies the UV condition, since in the IR $\vr >1$. The UV condition implies  the hierarchy 
\be
\label{moremore}
1 \ll \nf \ll \nc^{1/3} \,,
\ee
which is more stringent than \eq{stringent}. Similarly, the IR condition implies that 
\be
\vr \ll \vr_\mt{IR} \sac \vr_\mt{IR} = \frac{1}{\nf^2} \, \frac{\nc}{\nf}  \,,
\ee
which is more stringent than the IR part of \eq{low}. Note that \eq{moremore} is compatible with  \eq{stringent1} being more stringent than  \eq{stringent2} or vice versa. 

The second requirement for the DBI action to be valid is that the effective coupling between open strings be small. In the absence of smearing this coupling would be $e^{\phi} \nf$. However, in the presence of smearing not all the $\nf$ branes but only the fraction contained in a volume of string size can participate in a characteristic process involving open strings. As argued above, this fraction is $\nf \, \ell_s^2 / \ell^2$. The requirement that the effective string coupling (squared) be small is therefore 
\be
\frac{e^{2\phi} \, \nf^2  \, \ell_s^4}{\ell^4} \ll 1 \,.
\ee
Using the asymptotic expressions \eq{using} this equation turns into the two conditions
\bea
\mbox{IR:} \qquad && \frac{\nf}{\nc}\,  \frac{1}{\vr} \ll 1  \,, \\[2mm]
\mbox{UV:} \qquad && \frac{\nf}{\nc}\,  \frac{1}{\vr^2} \ll 1  \,.
\eea
The IR condition is automatically satisfied by virtue of \eq{moremore}, whereas the UV condition implies that
\be
\vr \gg \vr_\mt{UV} \sac \vr_\mt{UV} = \left( \frac{\nf}{\nc} \right)^{1/2}  \,,
\ee
which is less stringent than \eq{stringent1}. 

Putting together the various constraints coming from supergravity and from the DBI action we conclude that, in order for the gravity-plus-branes description to be valid, the number of colors and the number of flavors must obey \eq{moremore}. In this case the region of validity in terms of the $\vr$ coordinate is given by
\be
\label{low2}
\varrho_\mt{UV}  \ll \varrho \ll \varrho_\mt{IR} \sac 
\varrho_\mt{IR} = \frac{1}{\nf^2} \, \frac{\nc}{\nf}\sac 
\varrho_\mt{UV} = \max \left\{ \frac{1}{\nf} , \left( \frac{\nf}{\nc} \right)^{1/6} \right\} \,,
\ee
whereas in terms of the gauge theory coupling this region is characterized by  
\be
\gym^2 \ll 1 \sac 1 \ll \lambda \ll \min \left\{ 
\nc , \left( \frac{\nc}{\nf} \right)^{7/6} \right\} \,.
\ee
In Sec.~\ref{LPphysics} we will translate this region into a region in energy scale.

\section{Landau pole physics}
\label{LPphysics}
We will now examine some physical consequences of the presence of the Landau pole. 
The fact that the HV metric \eq{eq.LandaupoleasHV} has $\theta> p$ means that the asymptotic form of the metric satisfies the null-energy condition, as expected from the matter that sources it. More importantly, it implies  that the metric is conformal to AdS$_5$ with a conformal factor that vanishes faster than $u^{-2}$ as the Landau pole is approached. We will now see  that this property has far-reaching consequences, all of them consistent with the presence of a UV cut-off in the theory. 

\begin{figure}[t]
\begin{center}
\includegraphics[width=0.65\textwidth]{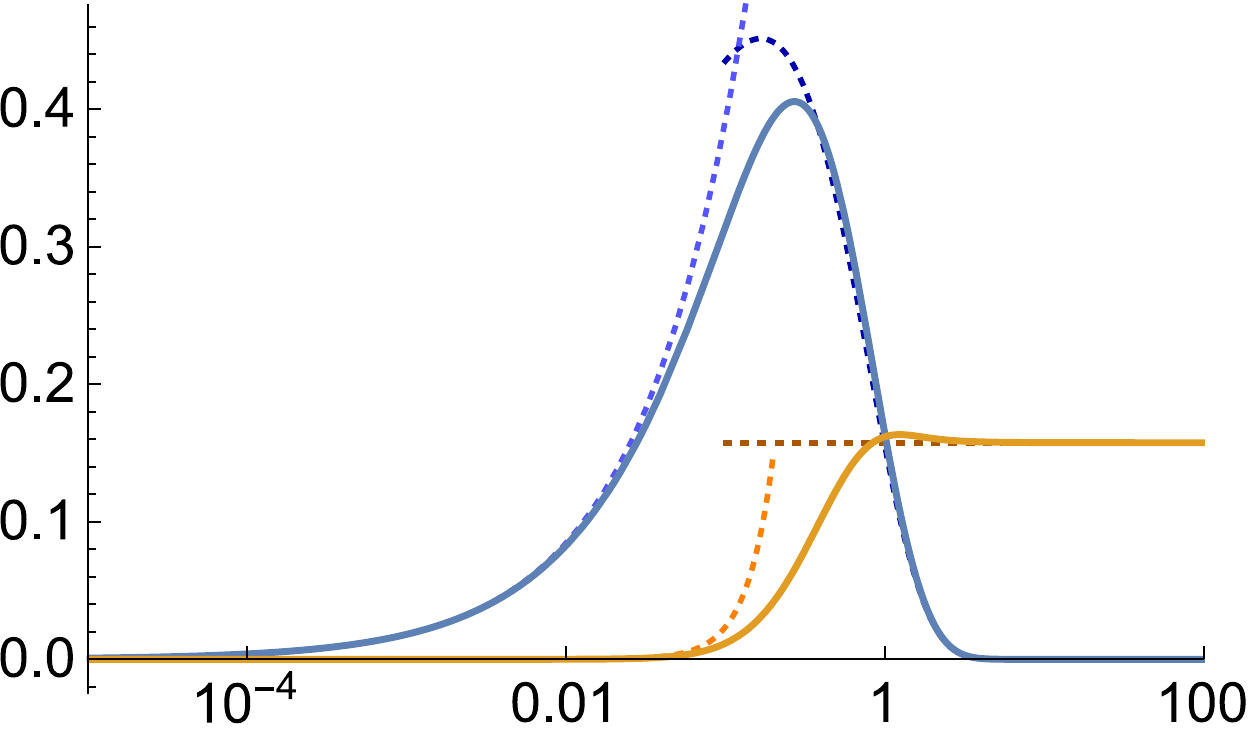} 
\put(-65,75){\Large $g_{\varrho\varrho}/\sqrt{\Qc}$}
\put(-220,50){\Large $|g_{tt}|=g_{xx}$}
\put(10,15){\Large $\varrho$}
\caption{Components of the five-dimensional metric \eqref{eq.susysolution2} for the supersymmetric solution. The dotted curves on the left- and the right-hand side indicate the UV  and the IR approximations to the solution given by \eq{eq.leadingUVsusy} and \eq{eq.leadingIRsusy}, respectively.}\label{fig.5dmetricsusy}
\end{center}
\end{figure}

A first consequence is that the components of the five-dimensional metric 
$|g_{tt}|=g_{xx}$ and $g_{\varrho\varrho} $ asymptotically decrease towards, and eventually vanish at, the Landau pole. Since these functions respectively vanish and go to a constant with positive corrections in the IR, it follows that they are non-monotonic functions of the holographic coordinate. This is shown in Fig.~\ref{fig.5dmetricsusy}, where we present both functions evaluated in terms of the radial coordinate $\varrho$ of \eqref{eq.susysolution2}, in which the IR is at $\varrho\to\infty$ and the LP is at $\varrho=\rholp\equiv 0$. In the figure we have also set $c_2=1$. 
This non-monotonicity has several implications. From a purely general-relativist's viewpoint, it signals a pathology since it means that gravity becomes repulsive in the UV region where $|g_{tt}|$ decreases. From a holographic viewpoint, it implies a minimum size and a maximum energy in the gauge theory associated to an object of fixed proper size and fixed proper energy in the bulk. Indeed, consider such an object in the IR part of the geometry. As we move this object closer to the boundary, its gauge theory size decreases and its gauge theory energy increases. This situation, which is the usual one in holography, underlies the correspondence between holographic position in the bulk and energy scale in the boundary \cite{Susskind:1998dq,Peet:1998wn}. However, beyond the turning point of 
$|g_{tt}|=g_{xx}$ this behavior is reversed and the gauge theory size of the object begins to increase while its gauge theory energy begins to decrease. Therefore, there is a minimum of the gauge theory size and a maximum of the gauge theory energy.  The existence of a minimum length scale or a maximum energy scale are both consistent with the presence of a UV cut-off in the gauge theory.

%JTJT
The non-monotonicity of the five-dimensional metric is intimately related to the behavior of the holographic $c$-function defined in \cite{Girardello:1998pd,Freedman:1999gp}. This function  satisfies two properties: its value coincides with the central charge of the corresponding conformal field theory at fixed points, and it is monotonically increasing towards the UV if the weak energy condition holds. For the D3-D7 solution \eqref{eq.susysolution2} the $c$-function was computed  in \cite{Bigazzi:2009gu}.\footnote{In \cite{Bigazzi:2009gu} the authors refer to the $a$-function since for theories dual to two-derivative gravity the central charges $a$ and $c$ coincide.} This reference found that the $c$-function takes the value 
$c=27 \nc^2/64$ at the deep IR, grows from there  towards the UV and diverges towards $+\infty$ at the point  where the components of the five-dimensional metric attain a maximum (see figure \ref{fig.5dmetricsusy}). Passed this point the $c$-function continues growing from $c=-\infty$ to $c=0^-$ at the Landau pole. 

While this result may suggest that the effective number of degrees of freedom   diverges at an energy scale below the Landau pole, one must remember that the $c$-function can be directly interpreted as the number of degrees of freedom only at a fixed point. Moreover, the thermal entropy that we will compute below actually shows a maximum number of degrees of freedom per unit volume. In any case, it would be interesting to investigate this issue further in the future.

\begin{figure}[t]
\begin{center}
\includegraphics[width=0.65\textwidth]{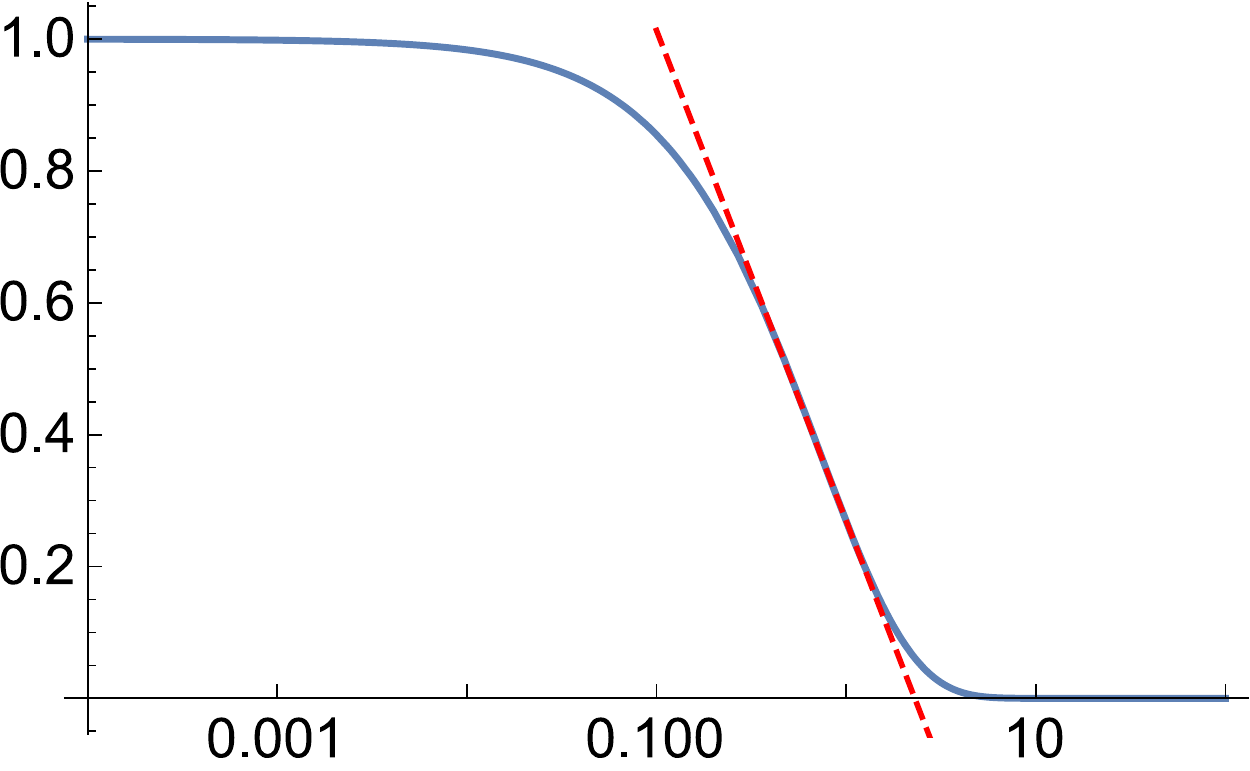}
\put(-300,190){\Large $ 
M(\varrho) / \lp$}
\put(10,15){\Large $\varrho$}
\caption{Mass of a  string stretching from the IR ($\varrho\to\infty$) to a radial distance $\varrho$. The dashed red line corresponds to $M = a_1 - a_2 \log \varrho$, with $a_1, a_2$ positive constants.}
\label{fig.stringlength}
\end{center}
\end{figure}

A second  consequence  is  that the proper distance along the radial direction to the end of the geometry is finite, since the integral
\be
\int^{u_\mt{UV}} \sqrt{g_{uu}}\, \d u \sim \int^{u_\mt{UV}} \frac{du}{u^{1+\frac{\theta}{p}}}  \sim 
\frac{1}{(u_\mt{UV})^\frac{\theta}{p} }
\ee
converges as $u_\mt{UV} \to \infty$. This should be contrasted with the pure AdS$_5$ case, in which $\theta=0$ and this distance diverges logarithmically. The fact that the Landau pole is located at a finite distance along the holographic direction is consistent with its interpretation as a maximum energy scale. This idea can be made more precise by noticing that there is a maximum mass for a string stretching  from the IR ($\varrho\to\infty$) to a brane at fixed radial position in the bulk. In the gauge theory this is interpreted as the self-energy of a charged particle, and it is given by
\be
M (\varrho)= \frac{1}{2\pi\ls^2} \int_{\varrho}^{\infty} 
\sqrt{-G_{tt}^\mt{st} \, G_{\varrho\varrho}^\mt{st}} \, \d \varrho 
\,,
\ee
where we recall that $G^\mt{st}$ is the ten-dimensional string-frame metric. Using the supersymmetric solution \eq{eq.susysolution2} (with $\rholp=0$) and \eq{eq.hsupersymmetric}  we find
\be
M (\varrho)= \frac{c_2}{2\pi \ell_s^2} \, \frac{\Qc^{1/4}}{\Qf^{1/2}} \,  (6e)^{1/6}\, \Gamma \left( \frac{2}{3} , \frac{1}{6} + \varrho \right) \,.
\ee
This result is plotted in Fig.~\ref{fig.stringlength}.
We see that the energy of the string attains the finite limit
\be
\label{mmax}
\lp= \frac{c_2}{2\pi\ls^2} \frac{\Qc^{1/4}}{\Qf^{1/2}} \, (6e)^{1/6}\, \Gamma \left( \frac{2}{3} , \frac{1}{6}  \right) 
\ee
as we integrate all the way to the end of the geometry at $\rholp=0$. This means that there is a maximum mass for an external quark that the gauge theory can be coupled to. As we have already indicated by our choice of notation in \eq{mmax} we will take this mass as the definition of the scale of the Landau pole  itself.  More generally, the function $M(\varrho)$ can be used to associate a gauge theory  energy scale  to a radial position on the gravity side in a diffeomorphism-invariant way. 
Note, however, that the absolute value of this energy is meaningless unless other scales are introduced in the theory. This is reflected in the fact that $\lp$ is proportional to the integration constant $c_2$, whose value cannot be fixed on physical grounds in the absence of other scales. Put differently, in the absence of other microscopic scales the Landau pole simply provides a unit of energy and can be set to any value.  In contrast, ratios of scales are meaningful, and in our case this is reflected in the fact that  the dependence on $c_2$ cancels out in such ratios. If one adds another scale to the theory, for example by introducing a non-zero temperature $T$ (as we will do in Sec.~\ref{app.neutralthermo}) or a non-zero quark mass $M_q$, then one can determine the absolute scale of the Landau pole, in units of $T$ or $M_q$, in terms of the value of the coupling measured at the scale $T$ or $M_q$. In our case the specification of the coupling at one of these physical scales would indeed fix $c_2$ and hence the overall normalization of $M(\varrho)$, thus making absolute scales and not just ratios of scales meaningful. 

\begin{figure}[t]
\begin{center}
\begin{subfigure}{.43\textwidth}
\includegraphics[width=\textwidth]{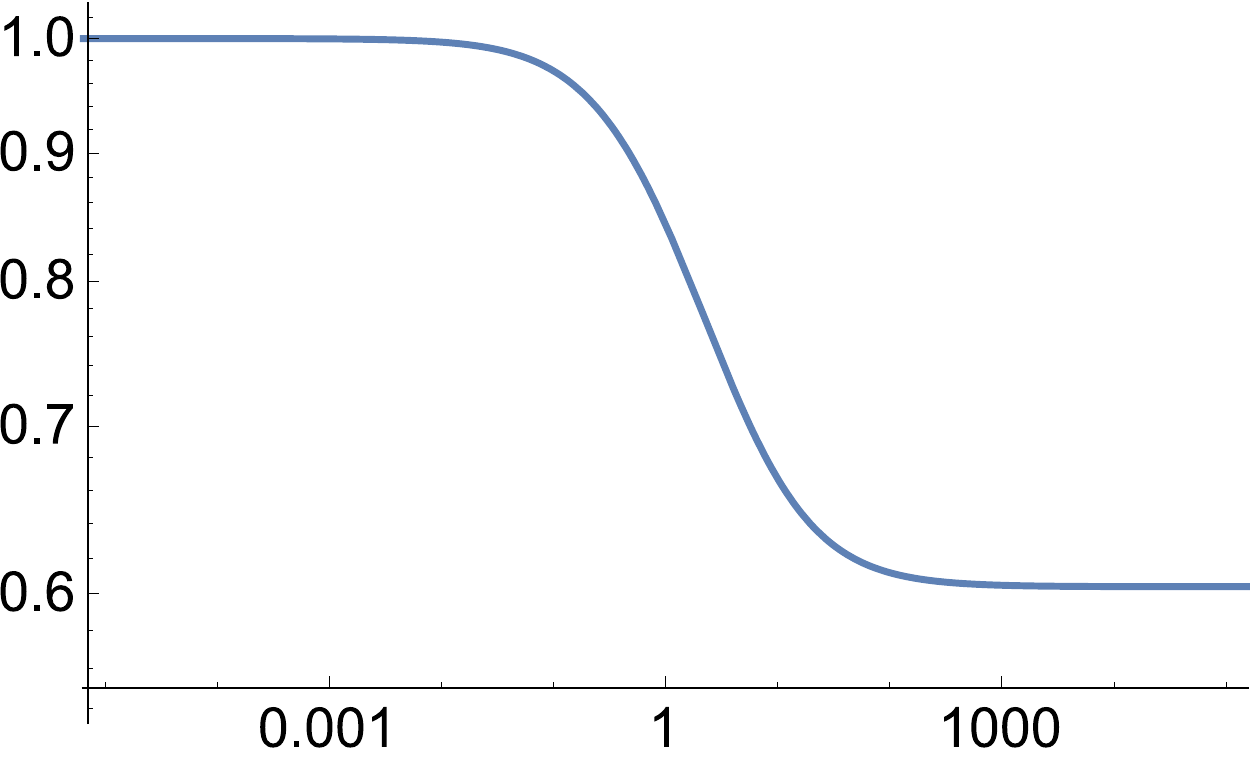} 
\put(-210,40){\rotatebox{90}{\large ${\beta(\gym^2)}/{\gym^4}$}}
\put(-100,-15){\large $\gym^2$}
\end{subfigure}\hspace{10mm}
\begin{subfigure}{.43\textwidth}
\includegraphics[width=\textwidth]{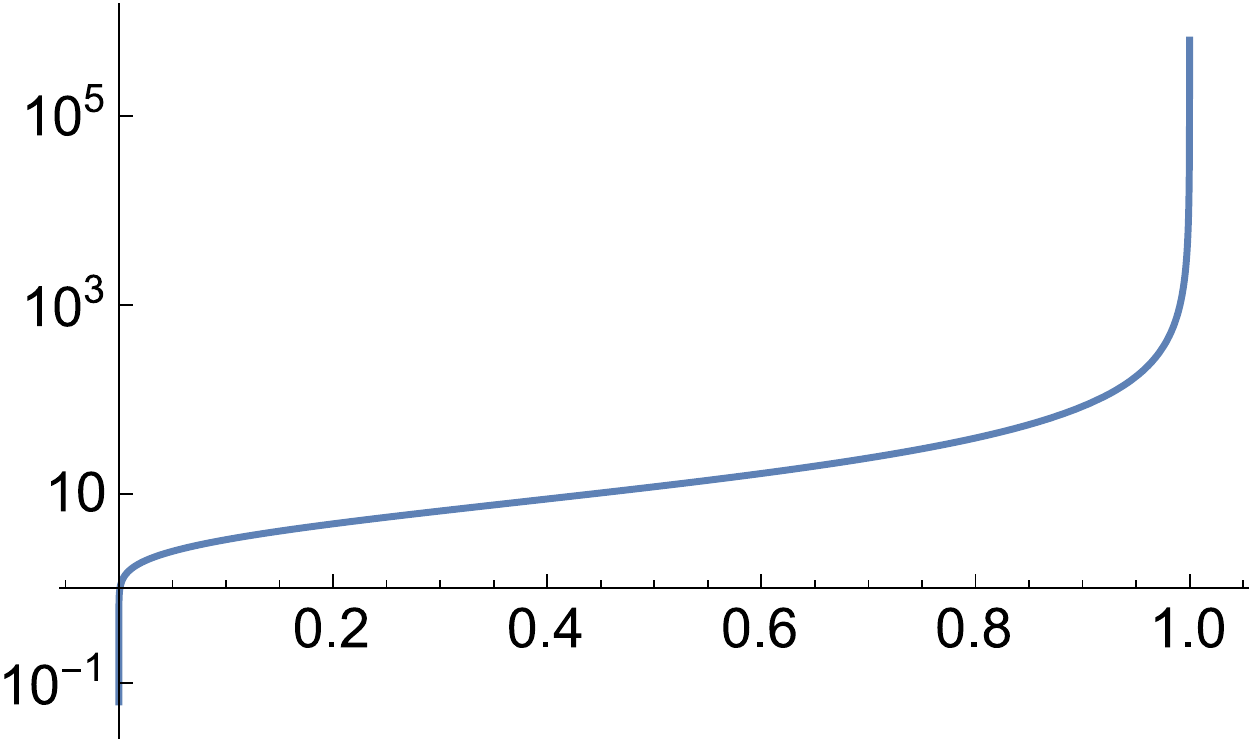} 
\put(-210,100){\large $\gym^2$}
\put(-100,-10){\large $M/\lp$}
\end{subfigure}\hspace{10mm}
\vspace{15mm}
\caption{(Left) Gauge theory $\beta$-function extracted from the supersymmetric gravity solution. We have chosen the overall normalization so that 
$\beta(\gym^2)/\gym^4 \to1$ as $\gym^2 \to 0$. With this choice $\beta(\gym^2)/\gym^4 \to e^{1/6}6^{-1/3}\Gamma[2/3,1/6] \sim 0.604$ as $\gym^2 \to \infty$. 
(Right) YM coupling as a function of the energy scale. 
}
 \label{fig.betafunction}
\end{center}
\end{figure}

With the map $M(\varrho)$ in hand we can  examine the dependence of the Yang-Mills coupling on the energy scale and extract the corresponding $\beta$-function as 
\be
\beta (\gym^2) = M\frac{d}{dM} \, \gym^2 = 
2\pi \, M\frac{d}{dM} \, e^{\phi(\rho(M))} \,.
\ee
We will ignore purely numerical factors and also the overall normalization of $M(\varrho)$. We begin in the IR. Expanding the $\Gamma$-function in the limit $\rho \to \infty$ we have
\vspace{1mm}
\be
\mbox{IR:} \qquad
M \sim e^{-\varrho} \qquad \to \qquad 
e^\phi \sim \frac{1}{\varrho} \sim - \frac{1}{\log M} \qquad \to \qquad  
\beta (\gym^2) \sim \gym^4 \,,
\ee 
where the first equations holds up to terms of  $O(\varrho^{1/3})$. In the UV we expand the $\Gamma$-function around $\vr=0$ to find 
\vspace{1mm}
\be
\mbox{UV:} \qquad
\lp-M \sim  \varrho \qquad \to \qquad 
e^\phi \sim \frac{1}{\varrho} \sim  \frac{1}{\lp- M} \qquad \to \qquad  
\beta (\gym^2) \sim \gym^4 \,.
\ee 
As shown by the dashed red line in Fig.~\ref{fig.stringlength} there also seems to be an intermediate regime where 
\be
\mbox{Interm.:} \qquad
M \sim - \log \varrho \qquad \to \qquad 
e^\phi \sim \frac{1}{\varrho} \sim e^M \qquad \to \qquad  
\beta (\gym^2) \sim \gym^2 \log \gym^2 \,.
\ee 
The full  $\beta$-function is shown in Fig.~\ref{fig.betafunction}(left), and the coupling itself in Fig.~\ref{fig.betafunction}(right).

It is interesting that the two different UV and IR asymptotic behaviors of $M(\varrho)$ give rise to two $\beta$-functions that, up to an overall constant, coincide with  the one-loop  $\beta$-function in a perturbative theory with positive $\beta$-function.

Using the UV and the IR forms of $M(\varrho)$ we see that the scales corresponding to the regime of validity of supergravity \eq{low2} satisfy
\be
M(\vr_\mt{UV}) \sim \lp \sac 
\frac{M(\vr_\mt{UV})}{M(\vr_\mt{IR})} \sim \exp \left( \frac{1}{\nf^2} \, \frac{\nc}{\nf} \right) \,.
\ee
The first of these equations says that the maximum energy scale in the region covered by supergravity is of the same order as the Landau pole scale. The second equation says that this scale is exponentially higher than the lowest IR scale in the regime described by \eq{moremore}. Note that the smaller $\nf$ is compared to $\nc$ the larger this range is. This hierarchy would have been even larger, of order $\exp (\nc/\nf)$, if we had only considered the constraints \eq{low} imposed by the validity of supergravity. In any case, this hierarchy is consistent with the general expectation based on the fact that the $\beta$-function is suppressed in the regimes \eq{low} or \eq{low2}, which implies  that the Landau pole must be pushed off to infinity if 
$\nc \to  \infty$ while  $\nf$ and the IR scale are kept fixed. 

A third  consequence is the existence of a maximum density of degrees of freedom per unit volume in the gauge theory, again as expected from the existence of a UV cut-off. This number can be obtained on the gravity side by computing the area of a constant-$t$, constant-$r$ surface in Planck units, following \cite{Susskind:1998dq}. The calculation can be done either in the five-dimensional metric \eq{eq.5dmetricdef} or in the ten-dimensional Einstein-frame metric \eq{eq.10dmetric}, including in the latter case the volume of the five-dimensional internal geometry. The result is of course the same. Expressed in terms of the five-dimensional quantities it is simply
\be
n \sim \ell_p^{-3} \sqrt{g_{xx}^3} \,.
\ee 
Since $g_{xx}$ attains a maximum as a function of the radial coordinate $\vr$, so does the density of degrees of freedom. In  the IR  this density grows with the energy scale $M$. However, this behavior is reversed if the energy increases  beyond the maximum of $g_{xx}$. As we will see below, the entropy density as a function of temperature, $s(T),$ behaves qualitatively in the same way as $n(M)$, which will imply a thermal  instability (negative specific heat) at temperatures above the scale where $g_{xx}$ attains its maximum.

%%%%%%%%%%%%%%%
%%%%%%%%%%%%%%%%%%%%%%%%%%%%%%%%
\section{Quark-antiquark potential and entanglement entropy}
\label{qqsec}
%%%%%%%%%%%%%%%%%%%%%%%%%%%%%%%%
%%%%%%%%%%%%%%%
 In this section we discuss two gauge-invariant observables, whose detailed calculations are given in Appendices \ref{wilson} and \ref{ent}. Our goal is to verify the expected effect of a UV cut-off in the theory. 
 
\begin{figure}[t]
\begin{center}
\begin{subfigure}{.45\textwidth}
\includegraphics[width=\textwidth]{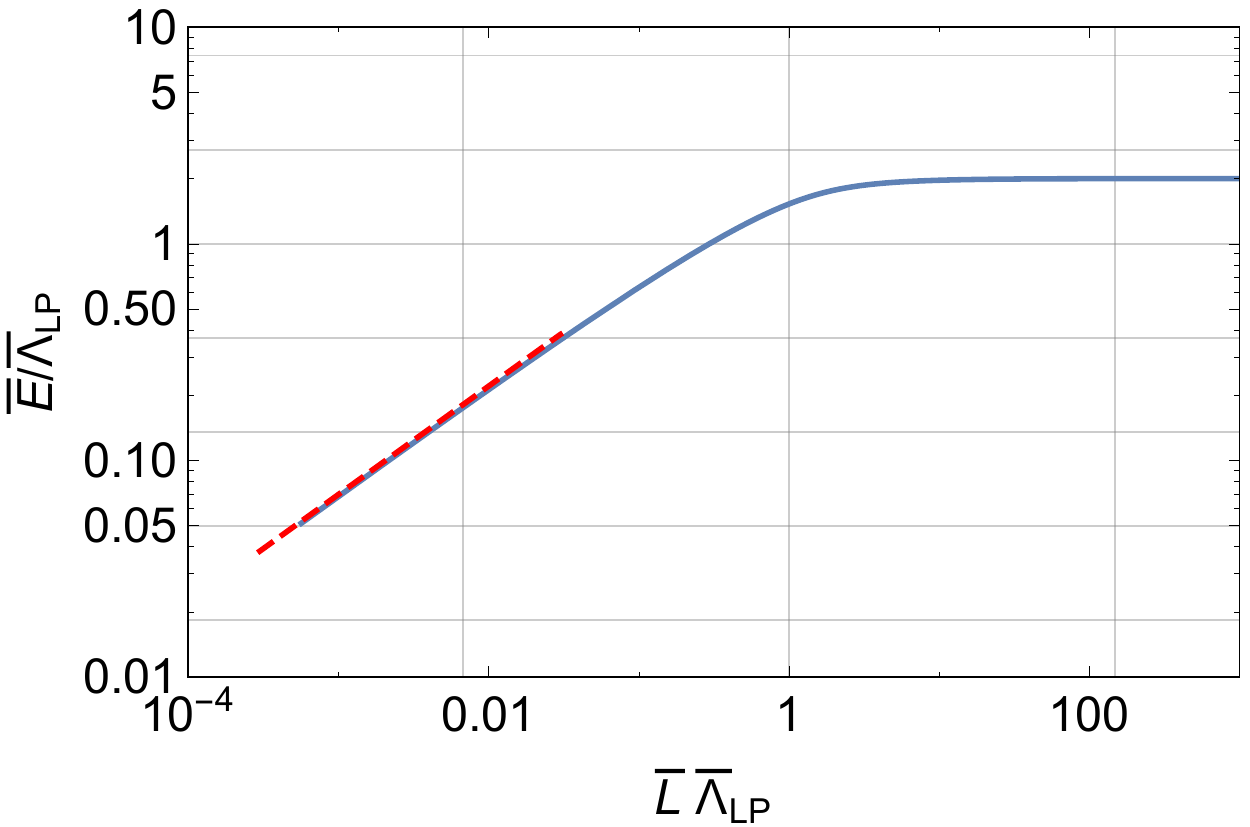} 
\end{subfigure}\hspace{10mm}
\begin{subfigure}{.45\textwidth}
\includegraphics[width=\textwidth]{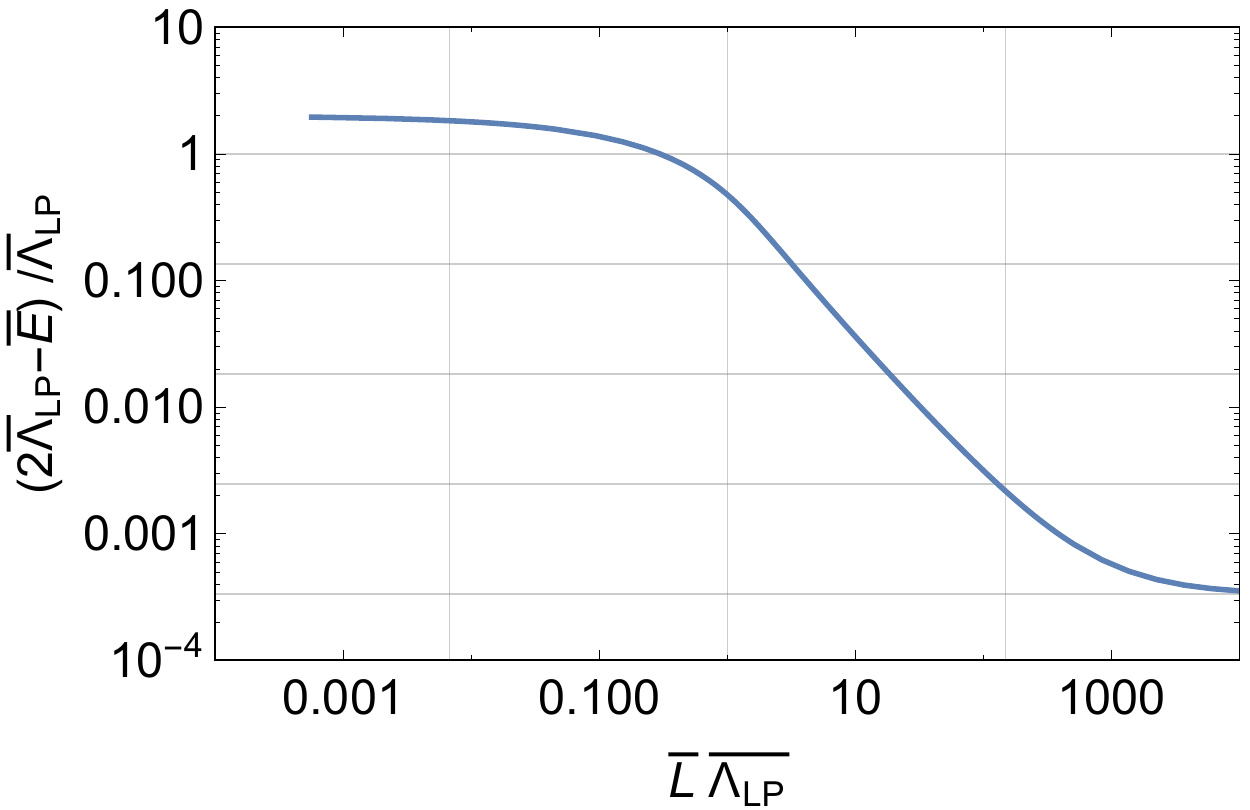} 
\end{subfigure}\hspace{10mm}
 \caption{Quark-antiquark potential for the supersymmetric solution. The dashed, red line on the left plot corresponds to the scaling $\bar E \sim \sqrt{\bar L}$. The dimensionless quantities 
 $\bar L$ and $\bar E$ are related to their dimensionful counterparts through Eqs.~\eq{redefL} and \eq{redefE}, respectively. The relation between $\lp$ and $\bar \Lambda_\mt{LP}$ is the same as that between $E$ and $\bar E$.
 }\label{fig:WLsusy}
\end{center}
\end{figure}

 The first observable is the Wilson loop associated to the worldlines of a static quark-antiquark pair separated a distance $L$ in the gauge theory.\footnote{The $L$ in this section should not be confused with that in \eqq{correct}.}  Note that in our theory this is not strictly speaking the same quantity as the quark-antiquark potential energy, since the string can break due to the presence of dynamical quarks (or, equivalently, D7-branes in the bulk). Nevertheless, for convenience we will still refer to the quantity extracted from the Wilson loop as the energy of a quark-antiquark pair. 
This quantity can be computed on the gravity side by considering a  fundamental string hanging form the quark and the antiquark. The result, given by \eqq{eee}  and plotted in \Fig{fig:WLsusy}, is finite without the need to subtract the energy of a disconnected quark-antiquark pair. 
In fact, not only the energy at arbitrary separations is finite, but also the short-distance limit, $L\to 0$, is finite. This limit is controlled by the near-Landau pole geometry and leads to the scaling $E \sim \sqrt{L}$, as indicated by the dashed, red line in \Fig{fig:WLsusy}(left). These properties are again manifestations of the finite UV cut-off set by the Landau-Pole. In the opposite limit, $L\to \infty$, the energy approaches a constant $E_\infty$ in such a way that  
\begin{equation}
(E-E_{\infty})\cdot L\sim (1-0.15\log_{10}L)\,.
\end{equation}  
In this limit the result is controlled by the IR, log-AdS geometry. Correspondingly, the pure AdS result  $(E-E_{\infty}) \sim 1/L$ receives a logarithmic correction.
In \Fig{fig:WLsusy}(right) we plot minus the binding energy of the quark-antiquark pair, defined by comparing its energy to that of two disconnected quarks: 
\be
- E_\mt{bind} = 2\lp - E \,.
\ee
We see from the plot that the binding energy is always negative, meaning that the connected configuration is always energetically preferred. We also note the curious fact that the binding energy does not approach zero as $L\to \infty$. In other words, there is a gap between the energy of the bound state and the energy of the disconnected quarks even when the separation between the constituents of the bound state is taken to infinity. This result should be taken with some caution, though, since for large $L$ the string explores the IR regime of the geometry and eventually becomes sensitive to the region where the supergravity+DBI description ceases to be valid. 

\begin{figure}[t]
\begin{center}
\begin{subfigure}{.48\textwidth}
\includegraphics[width=\textwidth]{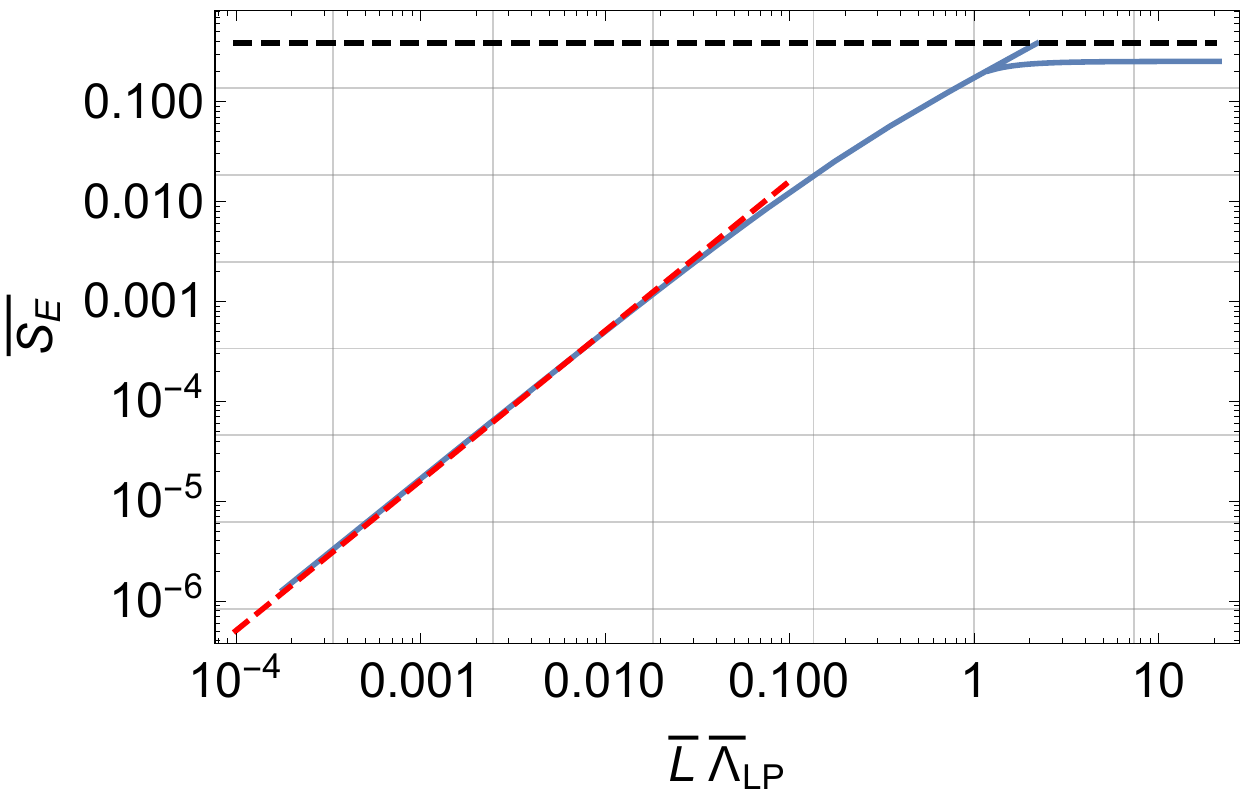} 
\end{subfigure}\hspace{5mm}
\begin{subfigure}{.48\textwidth}
\includegraphics[width=\textwidth]{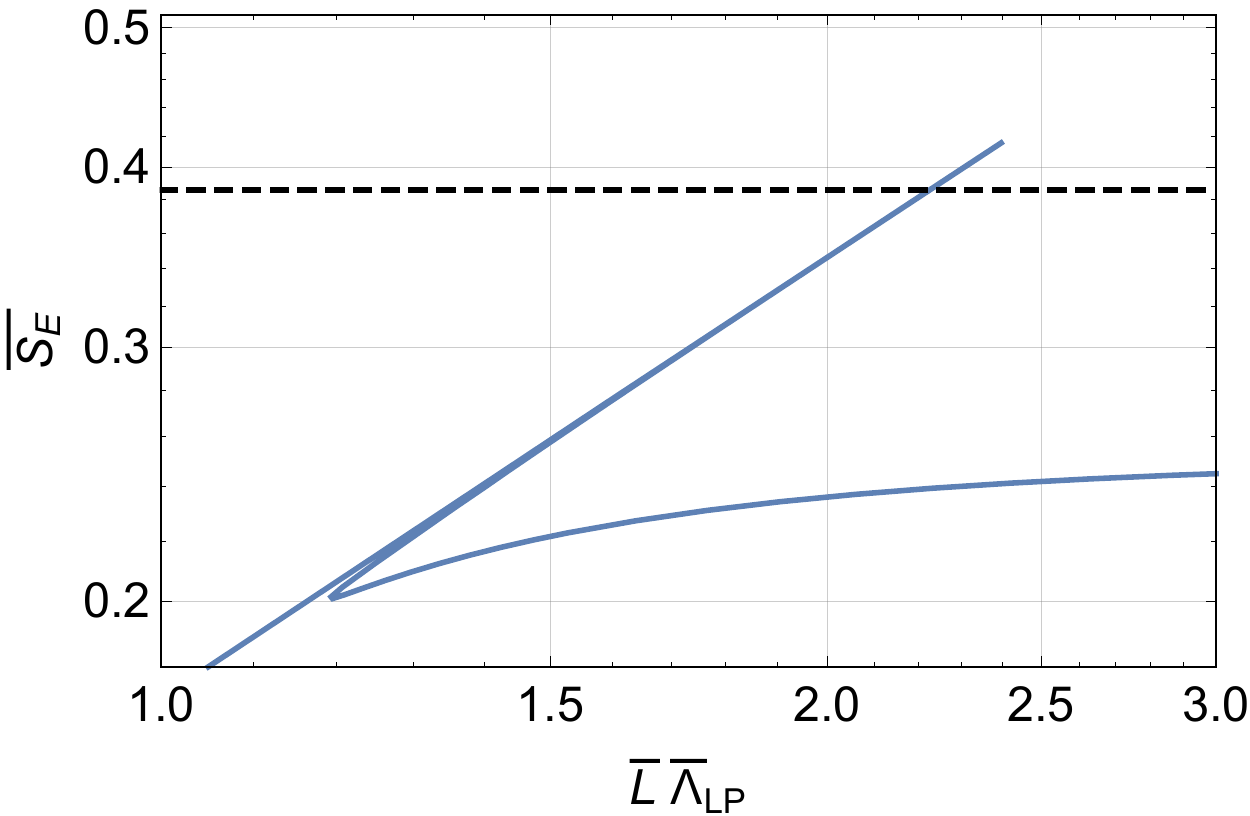} 
\end{subfigure}\hspace{10mm}
 \caption{Entanglement entropy, in the supersymmetric ground state of the gauge theory, for a three-dimensional stripe of width $L$. The dashed, red line on the left plot corresponds to the Landau pole scaling 
 $\bar S_E \sim \bar L^{3/2}$, whereas the dashed, black horizontal line is the EE of the disconnected configuration. The dimensionless quantities 
 $\bar L$ and $\bar S_E$ are related to their dimensionful counterparts through Eqs.~\eq{redefL} and \eq{redefS}, respectively. The relation between $\lp$ and $\bar \Lambda_\mt{LP}$ is the same as that between $E$ and $\bar E$ given in \eq{redefE}.}
 \label{fig:EEsusy}
\end{center}
\end{figure}

The second observable is the entanglement entropy (EE). In \Fig{fig:EEsusy} we show the result  for a three-dimensional stripe on the boundary as a function of its width $L$. For small values of the width we see the  scaling 
\be
\label{LPEE}
S_E\sim L^{3/2} \,, 
\ee
which is controlled by the Landau-Pole geometry. In the opposite limit, the entanglement entropy approaches a constant. The dashed, black, horizontal line in \Fig{fig:EEsusy}(left) is the EE of a disconnected configuration. These results exhibit one important similarity and one important difference with the quark-antiquark potential. The similarity is that the EE does not exhibit any UV divergence  (see e.g.~\cite{Nishioka:2009un}), again as expected for a theory with a UV cut-off. The difference is that the EE is multivalued in a certain range of widths. In this range there is more than one extremal surface in the bulk  anchored at the same region at the boundary. This leads to a phase transition in the EE at a value $\bar L \bar \Lambda_\mt{LP}\simeq 1.2$, as can be seen in \Fig{fig:EEsusy}(right). 
%DMDM
Note that, unlike in \cite{Klebanov:2007ws},  the transition occurs between two connected configurations, since the disconnected configuration never corresponds to that of minimal area. We will come back to this point in Sec.~\ref{disc}. 

%%%%%%%%%%%%%%%
%%%%%%%%%%%%%%%%%%%%%%%%%%%%%%%%
\section{Finite-temperature solutions}
\label{finiteT}
%%%%%%%%%%%%%%%%%%%%%%%%%%%%%%%%
%%%%%%%%%%%%%%%

We now turn to the construction of black hole solutions describing the finite-temperature states of the theory. In Ref.~\cite{Bigazzi:2009bk}  analytic, perturbative solutions in the flavor parameter $\Qf$  were constructed. The condition 
$\Qf\ll1$  implies a hierarchy between the scale of the Landau pole and the temperature, meaning that  the solutions constructed in \cite{Bigazzi:2009bk} are valid in the low-temperature regime, $T\ll \Lambda_{LP}$. Here we will go beyond this approximation by solving the equations of motion numerically. 

In order to integrate  the equations of motion for the functions given in the ansatz \eqref{eq.10dmetric} we choose a gauge for the radial coordinate such that the warp factor takes its AdS form
\be\label{eq.numericradial}
\sfh = \frac{\Qc}{4\, r^4} \ .
\ee
Making this choice, the constraint coming from the $rr$-component of Einstein's equations and the equation of motion for $\sfh$ allow us to solve algebraically for  $\sfc$ and  $\sfc'$, respectively. Substituting these back into the rest of the equations, we see that the function $\sfc$ drops out completely, in a fully consistent manner, and we are left with  four, second-order, ordinary differential equations (ODEs) for the functions $ \sff,\sfg,\phi$ and $\sfb$. However, one can further reduce this system of equations by requiring that
\be\label{eq.blackening}
\sfb = 1 - \frac{r_\mt{h}^4}{r^4} \ ,
\ee
with $r_\mt{h}$ the radius of the horizon. 
The equation of motion for $\sfb$ then becomes a first order ODE involving $ \sff,\sfg,\phi$ and it is a simple exercise to show that it is consistent with the 
second-order equations that these fields have to obey. All in all, the set of equations to be solved  consists of two second-order ODEs and one first-order ODE. The integration is done numerically, subject to appropriate boundary conditions, using a shooting method.

We now discuss the boundary conditions to be imposed. As $r\to \infty$ we demand that the solutions approach the Landau pole geometry. By comparing \eqref{eq.numericradial} to the UV behavior of the function $\sfh$ in the supersymmetric solution \eqref{eq.leadingUVsusy}, at leading order,  the relationship between the radial coordinates $\varrho$ and $r$ in the UV is given by (we set $c_2=1$)
\be
\mbox{UV}: \qquad \varrho = \frac{\Qc}{4\, r^4}  \,.
\ee 
In terms of the $r$ coordinate, the Landau pole geometry and the UV asymptotic expansion around it read
\be\label{eq.UVnumerics}
\bal
e^\sff & = \sqrt{\frac{3}{2}\, \Qc}\,\, 
\frac{1}{r^2} \Big[ 1 + \frac{\kappa_\sff}{r^4} +  \frac{\kappa_{\sff2}}{r^8} + {\cal O}(r^{-12}) \Big] \ , \\[2mm]
e^\sfg & = 1 - \frac{\kappa_\phi}{4r^4} + {\cal O}(r^{-8}) \ , \\[2mm]
e^\phi & = \frac{4\, r^4}{\Qf \, \Qc} \left[ 1 + \frac{\kappa_\phi}{r^4} + {\cal O}(r^{-8}) \right] \,,
\eal
\ee
where
\be
\kappa_{\sff2} = \frac{ 12\kappa_\sff^2 - 8 \kappa_\sff \kappa_\phi + \kappa_\phi^2 + 6 \Qc^2+ 8 \kappa_\sff \,r_\mt{h}^4  +4 \kappa_\phi\, r_\mt{h}^4}{ 24 } \, .
\ee
The expansion \eqref{eq.UVnumerics} is given in powers of $r^{-4}$  and is specified in terms of two unknown constants $\kappa_{\sff},\kappa_{\phi}$. Note that $\kappa_\sff$ and $\kappa_\phi$ have units of (length)$^4$ whereas $\kappa_{\sff2}$ has units of (length)$^8$. Since the only constant with units of length available in the setup is $\Qc^{1/4}$ it is convenient to work with the dimensionless quantities
\be\label{eq.scalings}
\frac{r}{\Qc^{1/4}} \ , \qquad \frac{\kappa_\sff}{\Qc} \ , \qquad \frac{\kappa_\phi}{\Qc} \ , \qquad \frac{\kappa_{\sff2}}{\Qc^2} \ .
\ee
Using these variables is equivalent to setting $\Qc=1$ in the equations of motion, i.e., using the D3-brane charge as the unit of length. Also, in the action \eqref{eq.neutral5daction}  the D7-brane charge, $\Qf$, always appears multiplying the dilaton, suggesting a  more appropriate variable
\be
\Qf \, e^\phi
\ee
in terms of which all the dependence on $\Qf$ has been modded out. As in the case above, this is equivalent to setting $\Qf=1$ in the equations of motion. In the numerical calculation we have  set $\Qc=\Qf=1$, but we present results here reinstating the appropriate factors of the charges.

In the IR we assume that we have a regular black hole horizon located at $r=r_\mt{h}$. We thus demand that,  as $r\to r_\mt{h}$, we can write a regular expansion of the form
\be\label{eq.IRparameters}
\bal
e^\sff  = e^{\sff_\mt{h}}+ {\cal O}(r-r_\mt{h}) \ , \\[2mm]
\quad e^\sfg = e^{\sfg_\mt{h}} + {\cal O}(r-r_\mt{h})  \ , \\[2mm]
e^\phi  = e^{\phi_\mt{h}} + {\cal O}(r-r_\mt{h}) \,.
\eal
\ee
This expansion is specified in terms of the radius of the horizon  $r_\mt{h}$ and three  dimensionless constants $\sff_\mt{h}, \sfg_\mt{h},\phi_\mt{h}$ that fix the values of the five-dimensional scalars at the horizon.

As explained above, the set of equations to be solved consists of two second order and one first order equations  and hence a solution is specified by five constants of integration. On the other hand, we have six parameters appearing in the boundary conditions. We thus expect to find a one-parameter family of solutions, labeled by $r_\mt{h}$,  for fixed $\Qc$ and $\Qf$.

\begin{figure}[t]
\begin{center}
\begin{subfigure}{.42\textwidth}
\includegraphics[width=\textwidth]{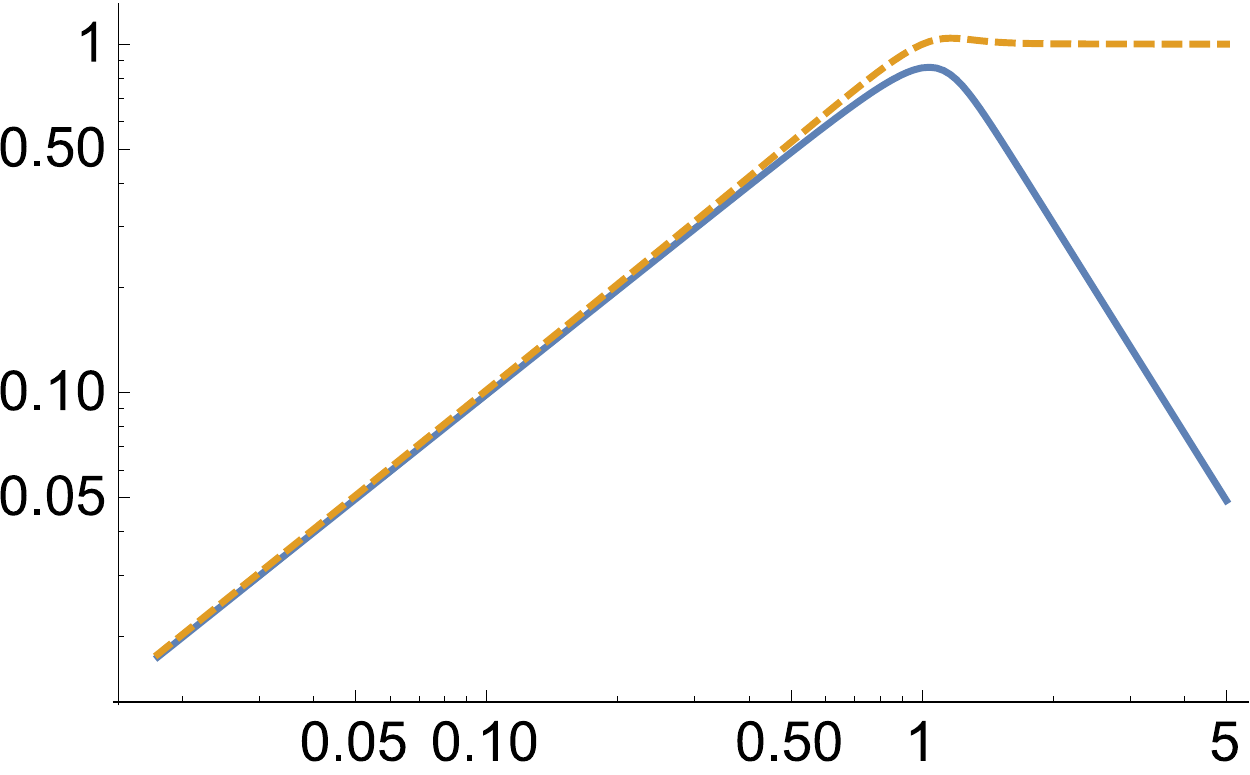} 
\put(-30,50){$e^{\sff_\mt{h}}$}
\put(-30,115){$e^{\sfg_\mt{h}}$}
\put(5,5){\Large$\frac{r_\mt{h}}{\Qc^{1/4}}$}
\end{subfigure}\hspace{10mm}
\begin{subfigure}{.42\textwidth}
\includegraphics[width=\textwidth]{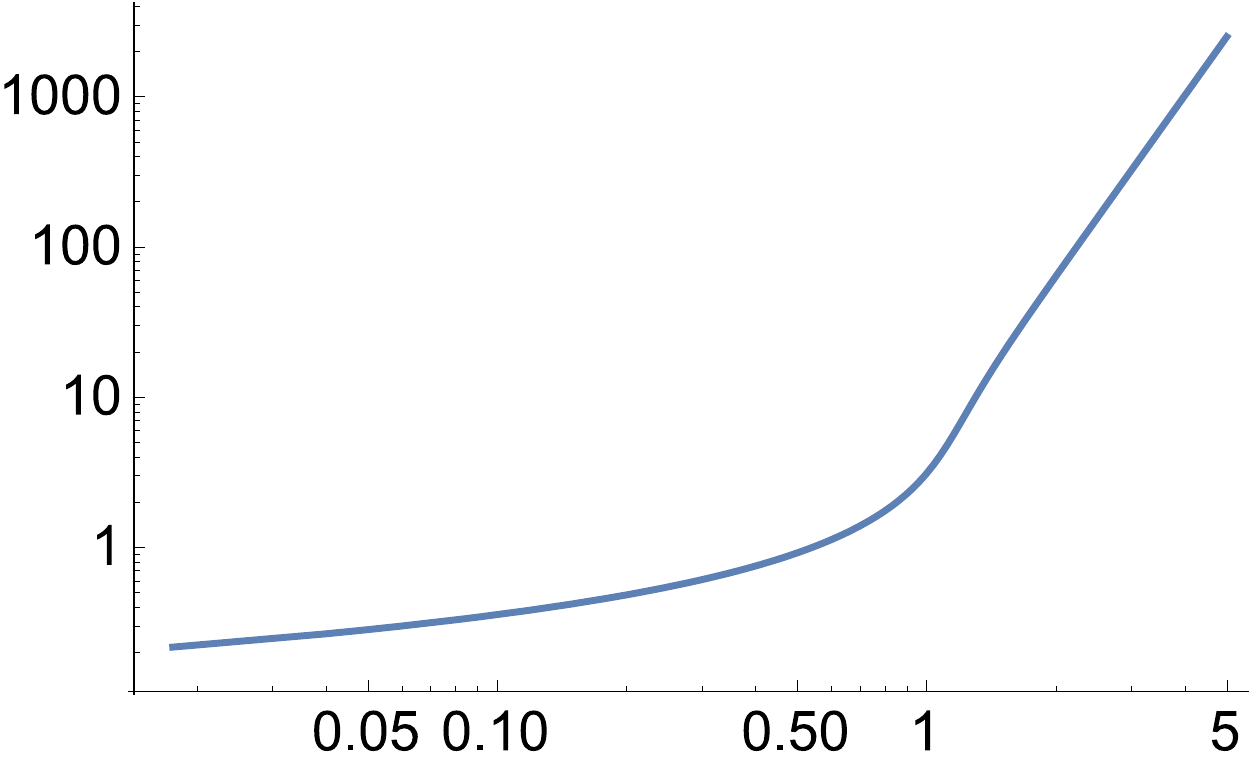} 
\put(-180,120){$\Qf \, e^{\phi_\mt{h}}$}
\put(5,5){\Large $\frac{r_\mt{h}}{\Qc^{1/4}}$}
\end{subfigure}\hspace{10mm}
\caption{IR coefficients of the  solutions as a function of the horizon radius.}\label{fig.neutralIR}
\end{center}
\end{figure}
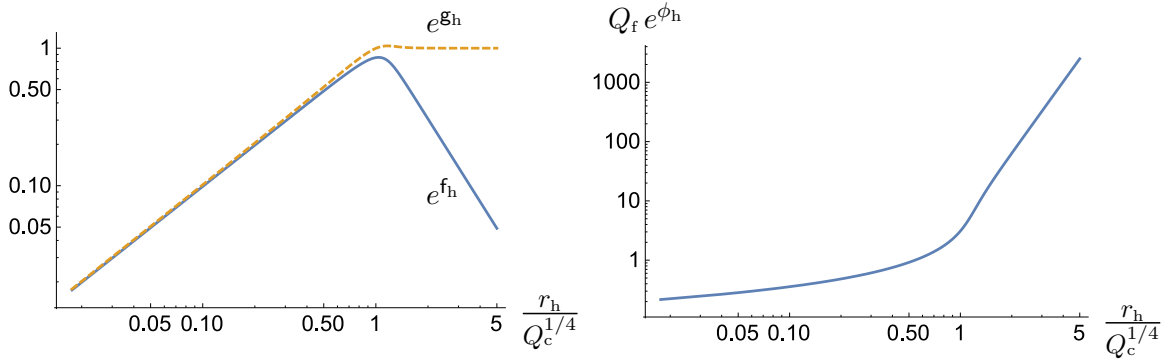

In \Fig{fig.neutralIR} we plot the numerically obtained parameters of the IR expansion \eqref{eq.IRparameters}. In these plots  we observe how the parameters nicely interpolate between  two different types of behavior, one for $r_\mt{h}\ll 1$ (cold solutions) and one for $r_\mt{h}\gg 1$ (hot solutions).  In the case of the hot solutions we see that the three coefficients follow power laws as functions of the horizon radius, seen as a line of constant slope on the logarithmic scale of the plot:  
\be
\mbox{Large $r_\mt{h}$:} \qquad e^{\sff_\mt{h}} = \sqrt{\frac{3}{2}}\, \frac{\Qc^{1/2}}{r_\mt{h}^{2}} \ , \qquad e^{\sfg_\mt{h}} = 1 \ , \qquad e^{\phi_\mt{h}} = \frac{4\, r_\mt{h}^4}{\Qf \,\Qc} \,.
\ee
These behaviors agree with those dictated by the leading terms in the UV expansion \eqref{eq.UVnumerics} if one cuts off the geometry at $r=r_\mt{h}$. In other words,  the IR coefficients at asymptotically high temperatures are  completely determined by the near-Landau pole geometry of the supersymmetric solution, since the blackening function \eqref{eq.blackening} increases rapidly from $\sfb=0$ at the horizon to $\sfb=1$, which is the supersymmetric value, when $r_\mt{h}/\Qc^{1/4}\gg 1$.

For the cold solutions we see that both $e^{\sff_\mt{h}}$ and $e^{\sfg_\mt{h}}$ follow again power laws as  functions of $r_\mt{h}$ given by
\be
\mbox{Small $r_\mt{h}$:} \qquad
e^{\sff_\mt{h}} =e^{\sfg_\mt{h}} = r_\mt{h}/\Qc^{1/4} \,,
\ee
with $e^{\sfg_\mt{h}}$ approaching this behavior with a positive correction and $e^{\sff_\mt{h}}$ approaching it with a negative correction.
This means that the squashing of the sphere persists in the near-horizon region, but this squashing decreases as the limit $r_\mt{h}\to 0$ is approached, in agreement with the zero-temperature, supersymmetric solution. The dilaton behavior is more intricate, and can be understood from the IR of the supersymmetric solution \eqref{eq.leadingIRsusy} with the change of radial coordinate 
\be\label{eq.IRchangeofradius}
\mbox{IR\,:} \qquad 
e^{-2\rho} \varrho^{1/3} = \frac{6^{-1/3}}{\sqrt{\Qc}} \, r^2 
\ee
applied to the supersymmetric dilaton profile
\be
e^{\phi_\mt{h}} = \frac{1}{\Qf \, \varrho_\mt{h}(r_\mt{h})}\ , 
\ee
where the function $ \varrho_\mt{h}(r_\mt{h})$ is given by the solution to \eqref{eq.IRchangeofradius}. All in all this shows that the IR parameters of the hot and cold solutions  are dictated by the UV and IR geometry of the supersymmetric solution, respectively. 
\begin{figure}[t]
\begin{center}
\begin{subfigure}{.42\textwidth}
\includegraphics[width=\textwidth]{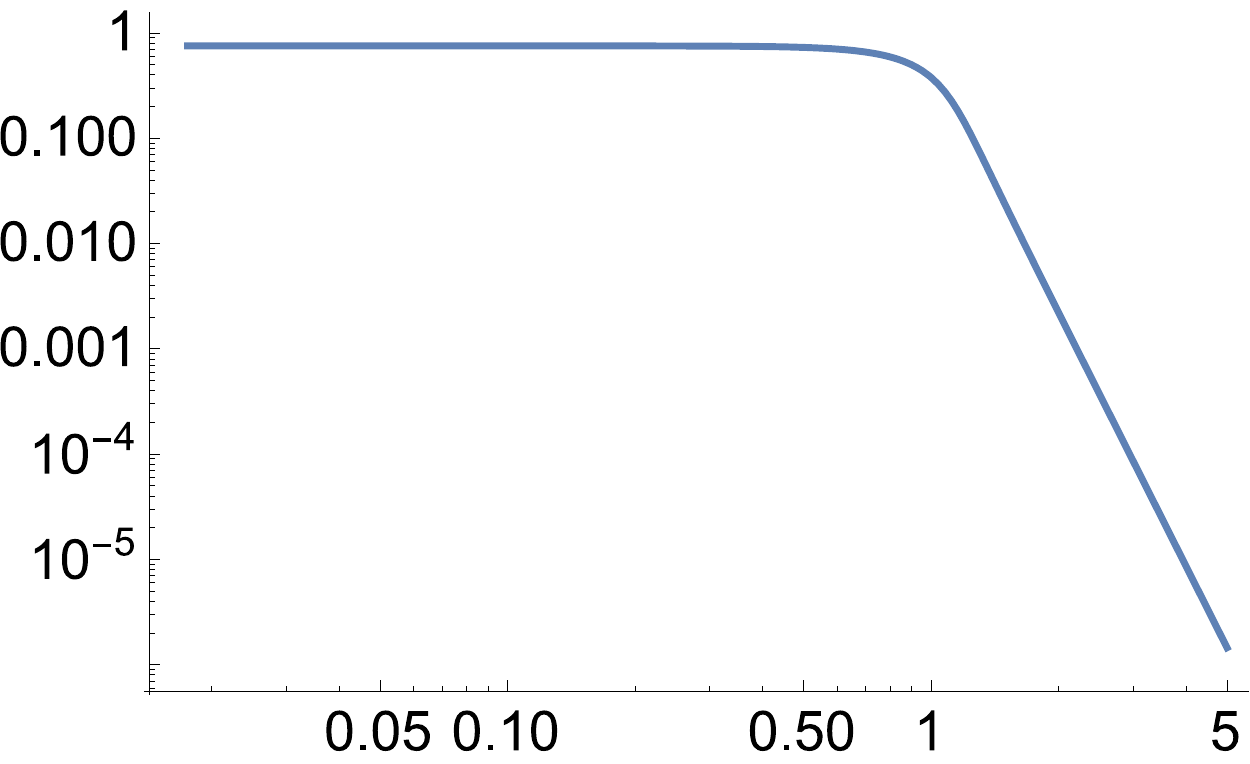} 
\put(-190,120){$-\kappa_\sff/\Qc$}
\put(5,5){\Large $\frac{r_\mt{h}}{\Qc^{1/4}}$}
\end{subfigure}\hspace{10mm}
\begin{subfigure}{.42\textwidth}
\includegraphics[width=\textwidth]{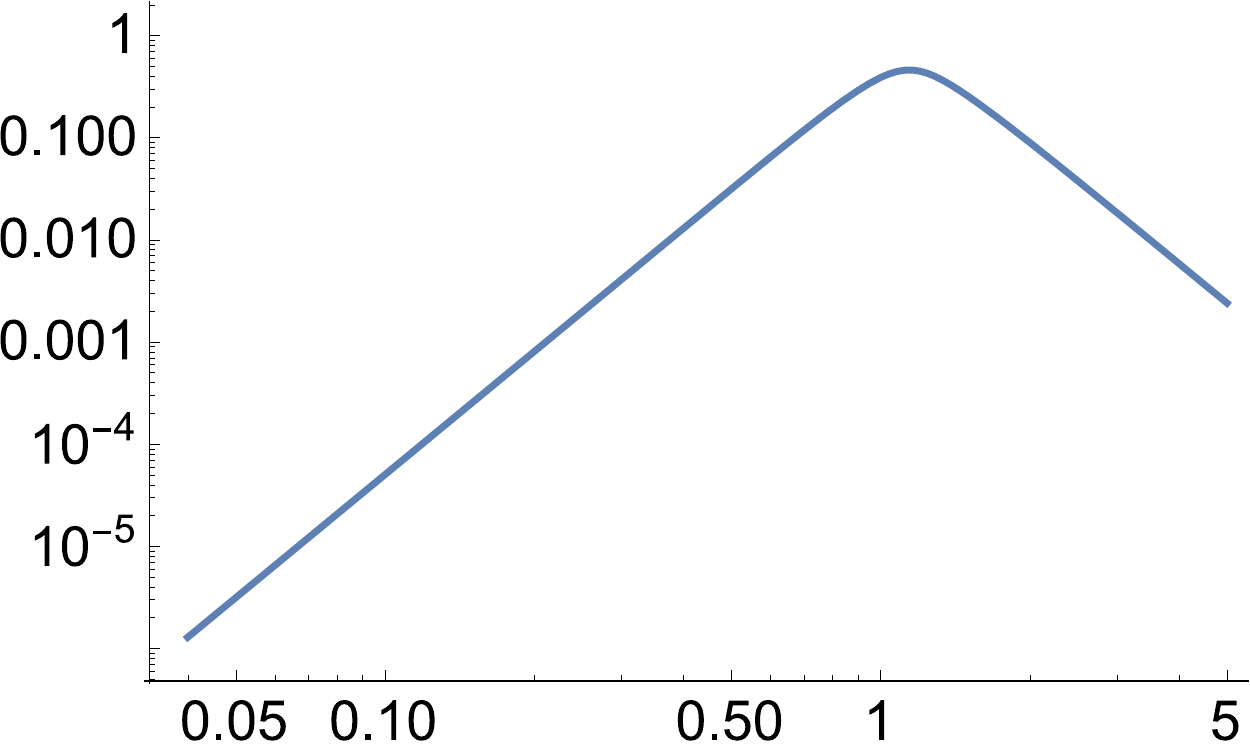} 
\put(-190,120){$-\kappa_\phi/\Qc$}
\put(5,5){\Large $\frac{r_\mt{h}}{\Qc^{1/4}}$}
\end{subfigure}\hspace{10mm}
\caption{UV coefficients as a function of the horizon radius.}
\label{fig.neutralUV}
\end{center}
\end{figure}

The UV coefficients are shown  in \Fig{fig.neutralUV}. 
These interpolate between two different behaviors for cold and hot solutions. The behaviors one obtains from the numeric solutions are
\be
\mbox{Small $r_\mt{h}$:} \qquad
\kappa_\sff = - \frac{3}{4}\Qc \ , 
\qquad \kappa_\phi = - \frac{1}{2}\, r_\mt{h}^4 \,, 
\ee
and
\be
\mbox{Large $r_\mt{h}$:} \qquad
\kappa_\sff = -0.55 \, \frac{\Qc^3}{r_\mt{h}^{8}} \ , 
\qquad \kappa_\phi = - \frac{3}{2}\, \frac{\Qc^2}{r_\mt{h}^{4}} \,. 
\ee
The first case corresponds to cold solutions and hence the value of the UV coefficients  can be obtained from those of the zero-temperature, supersymmetric solution. This is not possible for the large-$r_\mt{h}$ solutions.

%%%%%%%%%%%%%%%
%%%%%%%%%%%%%%%%%%%%%%%%%%%%%%%%
\section{Thermodynamics}\label{app.neutralthermo}
%%%%%%%%%%%%%%%%%%%%%%%%%%%%%%%%
%%%%%%%%%%%%%%%

The IR and UV parameters occurring  in the asymptotic expansions \eqref{eq.IRparameters} and \eqref{eq.UVnumerics} encode the thermal properties of the dual gauge theory.  The analysis of the thermodynamics requires the definition of an appropriate thermodynamic potential, in this case the free energy $F(T)$. In the holographic context this free energy is given by the renormalized on-shell action. The holographic renormalization procedure is well understood when the spacetime in question is asymptotically (locally) AdS, corresponding to a UV fixed point in the gauge theory. Nevertheless, we show in Appendix~\ref{continuation} that holographic renormalization can also be fully implemented in our case despite the presence of a singularity at a finite value of the holographic coordinate. The key observation is that the HV metric \eq{eq.LandaupoleasHV} can be related to an AdS metric through analytic continuation in the number of dimensions, along the lines of \cite{Kanitscheider:2009as,Gouteraux:2011ce}. Thus, in order to holographically renormalize our Landau pole geometry we simply need to analytically continue back the results from the AdS case.  
%DMDM
We emphasize that this procedure does not provide a UV completion of the theory but simply a prescription to renormalize a UV-incomplete theory. As we will now see, the result is quite intuitive.
%DMDM
\footnote{An alternative way to proceed would be to use the results of \cite{Chemissany:2014xsa}, where the procedure of holographic renormalization was worked out for asymptotically HV Lifshitz metrics. Our solution falls in the class labelled IIIb in Table (1.8) of \cite{Chemissany:2014xsa}.}

Our starting point is the Euclidean version of the bulk action \eqref{eq.neutral5daction}. Since the Einstein equations imply that the on-shell Lagrangian can be expressed as a total derivative, the on-shell action can be written as
\be
\label{diverges}
I_\mt{on-shell} = \frac{1}{2\kappa_5^2} \int \d^4 x  \int\d r \, 2\, \sqrt{g}\, 
R_{\,x}{^x} = - \frac{\beta\, V}{2\kappa_5^2} \frac{\sqrt{g_{tt}\, g_{xx}} \,g_{xx}'}{\sqrt{g_{rr}}} \Bigg|_{r \to \infty} \ ,
\ee
where the contribution at the horizon vanishes, $V$ is the spatial three-volume and $\beta=1/T$ is the period of the Euclidean time. Despite the fact that the Landau pole at $r=\infty$ is at a finite proper distance, \eq{diverges} diverges as $r^4$ due to the singularity at the Landau pole. Remarkably, this divergence is exactly cancelled by adding the standard boundary terms consisting of the Gibbons-Hawking term plus the usual counterterms for an AdS gravity-plus-scalars type of action. These are given by  
%DMDM
(see e.g. \cite{Papadimitriou:2004rz,Batrachenko:2004fd})
\be
\label{count}
I_\mt{boundary} = \frac{1}{2\kappa_5^2} \int \d^4 x \, \sqrt{\gamma} \left( 2K+{W} \right)  \Big|_{r \to \infty} \,,
\ee
with $\gamma_{ab}$ the induced metric on constant-$r$ slices, $K=K^a{_a}$ the trace of the extrinsic curvature of these slices, and $W$ the superpotential \eqref{eq.superpotential}. This contribution also diverges as $r^4$, but exactly with the right coefficient to cancel out the divergence present in $I_\mt{on-shell}$, thus giving a finite result for the total  action\footnote{These counterterms have been used in the same system already in Ref.~\cite{Bigazzi:2009tc}, where transport coefficients were studied.}
\be
I_\mt{renormalized} = I_\mt{on-shell} + I_\mt{boundary} = - \frac{4 \beta\,  V}{L^5\kappa_5^2} \left(  \kappa_\sff -  \kappa_\phi - r_\mt{h}^4 + 3 L^4 \right) \,.
\ee
The fact that the counterterms \eq{count} take the familiar form is the intuitive result that we referred to at the beginning of this section, and it follows from the analytic continuation discussed in Appendix \ref{continuation}.

The free energy density of the theory is given by 
\be
F=- I_\mt{renormalized}/\beta V \,.
\ee
The boundary stress-energy tensor also follows straightforwardly from the functional derivative of the renormalized on-shell action with respect to the metric induced on a constant-$r$ surface. The result takes the standard form 
\be
T^{\,a}{_b} = \frac{V}{2\kappa_5^2} \sqrt{\gamma} 
\Big[ -2 K^a{_b} + 2 \delta^a{_b} \left( 2K+ { W} \right) \Big]_{r\to\infty} = \mathrm{diag} \left(- E,P,P,P\right) \,,
\ee
where the explicit expressions for the energy and the pressure in our case are
\be
E = 2\, \frac{2 \kappa_\sff - 2 \kappa_\phi - r_\mt{h}^4 + 6L^4}{L^5\, \kappa_5^2} \ , \qquad P = - F \,. 
\ee
We can now perform a check of our calculations. On the one hand the expressions above determine the product $TS$ as
\be
TS = E-F \,.
\ee
On the other hand this product can be computed directly from the horizon expressions for $T$ and $S$: 
 \be
T = \frac{ r_\mt{h}^{6}}{4\sqrt{2}\, e^{\sff_\mt{h}+4\sfg_\mt{h}} L^7 \pi} \ , \qquad S = \frac{8\sqrt{2}\, e^{\sff_\mt{h}+4\sfg_\mt{h}} L^2 \pi}{ r_\mt{h}^2 \kappa_5^2} \ .
\ee
It is easy to check that both calculations yield the same result. We have also verified the  first law of thermodynamics in differential form
\be
S = - \frac{\d F}{\d T} \,,
\ee
by computing both sides in terms of the coefficients shown in Figs.~\ref{fig.neutralIR} and \ref{fig.neutralUV}.

From now on we will factor out appropriate powers of the radius $L$ in order to work with dimensionless quantities. For example, we will work with a dimensionless free energy $\mathcal{F}$ defined through
\be
F = \frac{4}{L^5\kappa_5^2} \left(  \kappa_\sff -  \kappa_\phi - r_\mt{h}^4 + 3 L^4 \right) = \frac{16}{L\, \kappa_5^2} \left( \frac{\kappa_\sff}{\Qc} -  \frac{\kappa_\phi}{\Qc} - \frac{r_\mt{h}^4}{\Qc} + \frac{3}{4}  \right) \equiv 
\frac{L^3}{\kappa_5^2}  \, \frac{\cal F}{L^4} \ .
\ee
The analogous expressions for the energy, the temperature and the entropy are 
\be\bal
E & = \frac{16}{L\, \kappa_5^2} \left( \frac{\kappa_\sff }{\Qc}-  \frac{\kappa_\phi}{\Qc} - \frac{r_\mt{h}^4 }{2\Qc}+ \frac{3}{4} \right) \equiv  
\frac{L^3}{\kappa_5^2} \,\frac{\cal E}{L^4} \ , \\[3mm]
T & = \frac{\sqrt{2}\, r_\mt{h}^6/\Qc^{3/2}}{e^{\sff_\mt{h}+4\sfg_\mt{h}} \pi \,  L} \equiv  \frac{\cal T}{ L} \ , \\[3mm]
S & = \frac{8\sqrt{2}\, e^{\sff_\mt{h}+4\sfg_\mt{h}} L^2 \pi}{ r_\mt{h}^2 \kappa_5^2} \equiv \frac{L^3}{\kappa_5^2} \, \frac{\cal S}{L^3} \,.
\eal\ee
Recalling Eqs.~\eq{correct} and \eq{kkappa2} we see that $F$, $E$ and $S$ scale as 
$\nc^2$, as expected, and that we are simply using $L$ as a unit with which to measure all dimensionful quantities. The dimensionless functions above obey the same relations as their dimensionful counterparts, namely
\be
{\cal F} = {\cal E} - {\cal T} {\cal S} \ , \qquad \d {\cal F} = - {\cal S} \, \d {\cal T} \ .
\ee
For later reference we also introduce a dimensionless specific heat 
\be
\label{heat}
C_V=T\frac{\d S}{\d T} = \frac{1}{\kappa_5^2} {\cal T} \frac{\d {\cal S}}{\d {\cal T}} \equiv \frac{L^3}{\kappa_5^2} \, \frac{\cal C_V}{L^3} \,, 
\ee
in terms of which the speed of sound may be written as 
\be
\label{sound}
v_s^2 = \frac{\d P}{\d E} = \frac{\d \cal P}{\d \cal E} = \frac{\cal S}{\cal C_V} \,.
\ee

\begin{figure}[t]
\begin{center}
\begin{subfigure}{.42\textwidth}
\includegraphics[width=\textwidth]{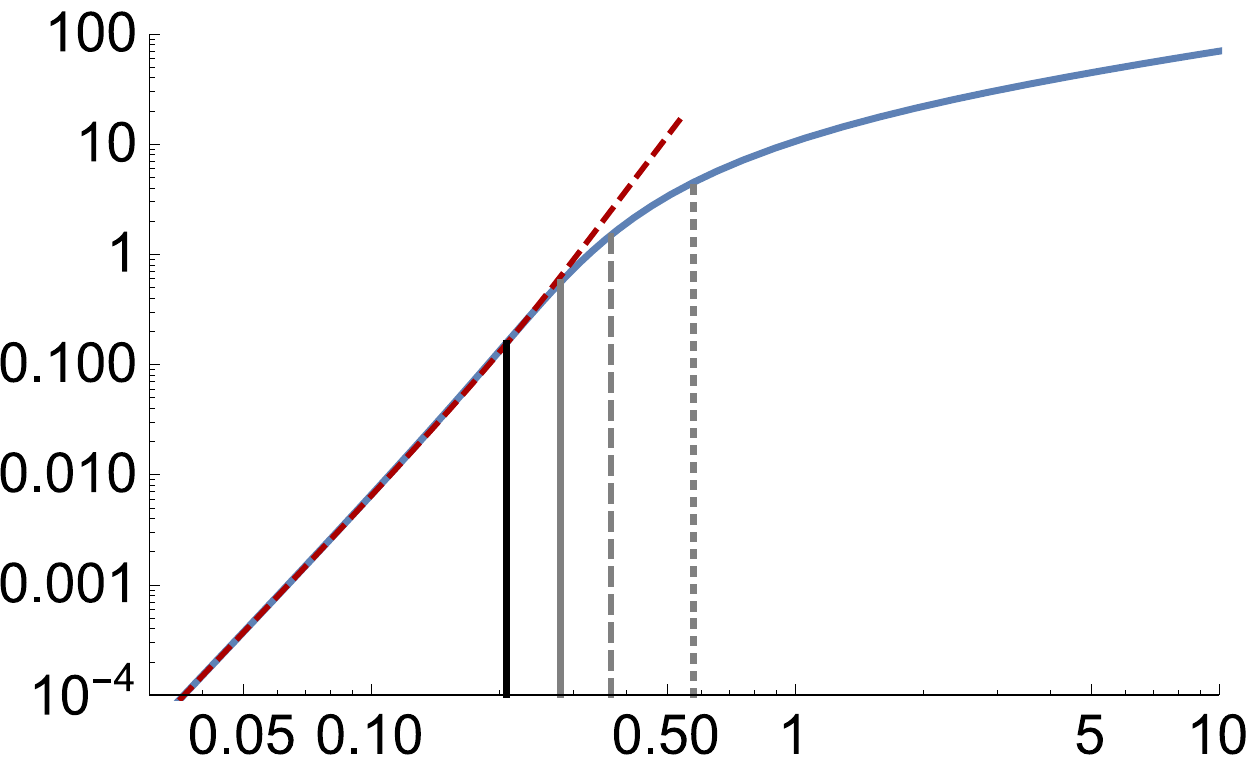} 
\put(-190,120){\large $-\cal F$}
\put(5,5){\large $\cal T$}
\end{subfigure}\hspace{10mm}
\begin{subfigure}{.42\textwidth}
\includegraphics[width=\textwidth]{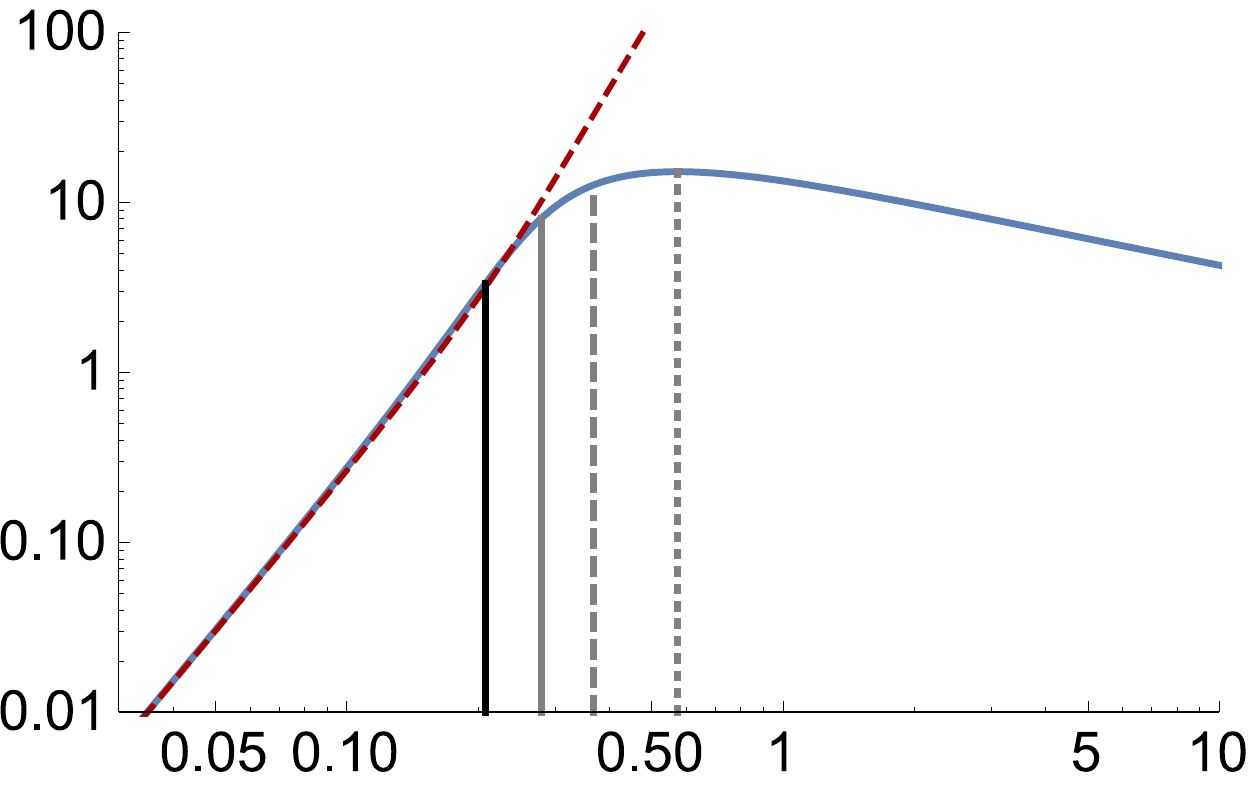} 
\put(-180,120){\large $\cal S$}
\put(5,5){\large $\cal T$}
\end{subfigure}
\caption{Log-log plot of the dimensionless free energy and entropy densities as functions of the dimensionless temperature. The dashed, red curves correspond to the low-temperature approximation \eq{eq.d3d7qgpresult}. The vertical lines correspond to the different temperatures defined around and below \eq{around}.} \label{fig.thermo}
\end{center}
\end{figure}

\begin{figure}[t]
\begin{center}
\includegraphics[width=.6\textwidth]{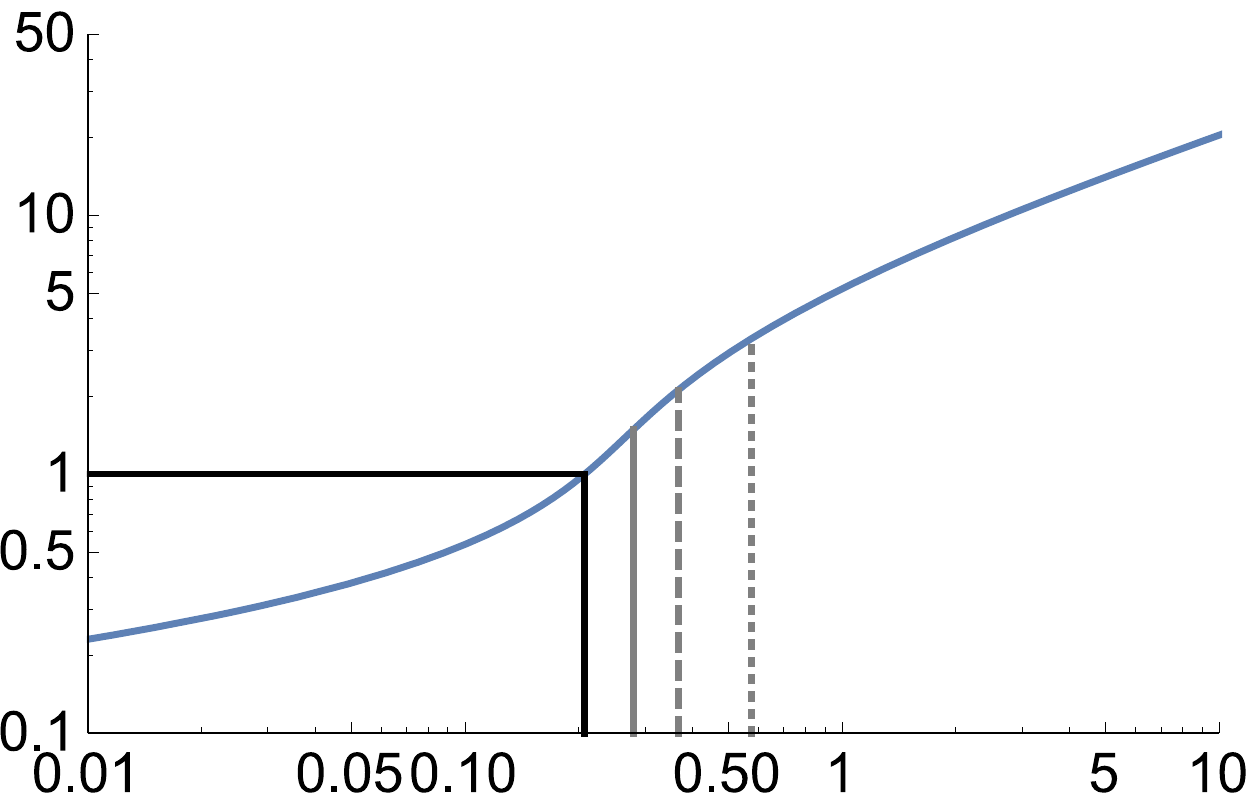} 
\put(-275,185){\Large $\Qf \, e^{\phi_\mt{h}}$}
\put(10,12){\Large $\cal T$}
\caption{Log-log plot of the horizon value of the dilaton as a function of the dimensionless temperature.}\label{fig.dilvsT}
\end{center}
\end{figure}

Substituting the coefficients shown in Figs.~\ref{fig.neutralIR} and \ref{fig.neutralUV} in the expressions above we obtain all the thermodynamic quantities. The free energy and the entropy densities are plotted  as a function of the temperature in Fig.~\ref{fig.thermo}, and the horizon value of the dilaton is plotted in  \Fig{fig.dilvsT}.
As mentioned above, Ref.~\cite{Bigazzi:2009bk} constructed black hole solutions in a perturbative expansion in $\Qf \,e^{\phi_\mt{h}}\sim \nf \, \gym^2 (T)$ that are valid  in the limit $T\ll \Lambda_\mt{LP}$. Their result for $\mathcal{F}$ and $\mathcal{S}$ to quadratic order in the expansion parameter is\footnote{Ref.~\cite{Bigazzi:2009bk} renormalized the free energy by subtracting the zero-temperature, supersymmetric solution, instead of via the inclusion of counterterms, as we have done here. Nevertheless, the fact that both prescriptions preserve supersymmetry at $T=0$ (which implies that $F$ vanishes in this limit) guarantees that both prescriptions yield the same result.}
\be
\label{eq.d3d7qgpresult}
\bal
{\cal F} & = - \frac{\pi^4}{2} {\cal T}^4 \left[ 1 + \frac{1}{2} \Qf \,e^{\phi_\mt{h}({\cal T})}  + \frac{1}{6} \left( \Qf\, e^{\phi_\mt{h}({\cal T})} \right)^2 + \cdots \right]\ , \\[2mm]
{\cal S} & = 2\pi^4 {\cal T}^3 \left[ 1 + \frac{1}{2} \Qf \,e^{\phi_\mt{h}({\cal T})}  + \frac{7}{24} \left( \Qf \,e^{\phi_\mt{h}({\cal T})} \right)^2 + \cdots \right]\,.
\eal\ee
Note that the leading term is precisely the same as in a CFT (which is $\mathcal{N}=4$ SYM if the internal manifold is S$^5$). This illustrates  the fact that, in the IR, the log terms in the metric \eq{IRmetric} are a small correction to the AdS$_5$ metric. 
We see in our dilaton plot that, indeed, $\Qf \,e^{\phi_\mt{h}}\ll1$ at low temperatures. Using this dilaton profile we have checked that our numerical results for $\mathcal{F}$ and $\mathcal{S}$  agree with \eq{eq.d3d7qgpresult} in the region ${\cal T}\ll 1$, as can be clearly seen in Fig.~\ref{fig.thermo}. 

The behavior at high temperatures is dramatically different. In this region we find 
%DMDM
\be
{\cal F} \sim - {\cal T}^{\,d_\mt{eff}} 
 \sac {\cal S} \sim {\cal T}^{\,d_\mt{eff}-1} \,,
\ee
where 
\be
d_\mt{eff} \,=\, p-\theta+1 \,=\, \frac{1}{2}\,.
\label{deffmain}
\ee
This  power-law behavior is  dictated by the asymptotic form of the HV metric \eqref{eq.LandaupoleasHV} with $p=3$ and  $\theta=7/2$. In other words,  the high-temperature thermodynamics is determined by the Landau pole. We see that the thermodynamic quantities in the UV effectively scale as in a \emph{lower}-dimensional conformal theory of dimension $d_\mt{eff} < 4$. This is consistent with the fact that the analytic continuation needed to define holographic renormalization for the HV metric connects this metric to an AdS metric in $d_\mt{eff}+1$ dimensions, see \eqq{deff}.

\begin{figure}[t]
\begin{center}
\begin{subfigure}{.42\textwidth}
\includegraphics[width=\textwidth]{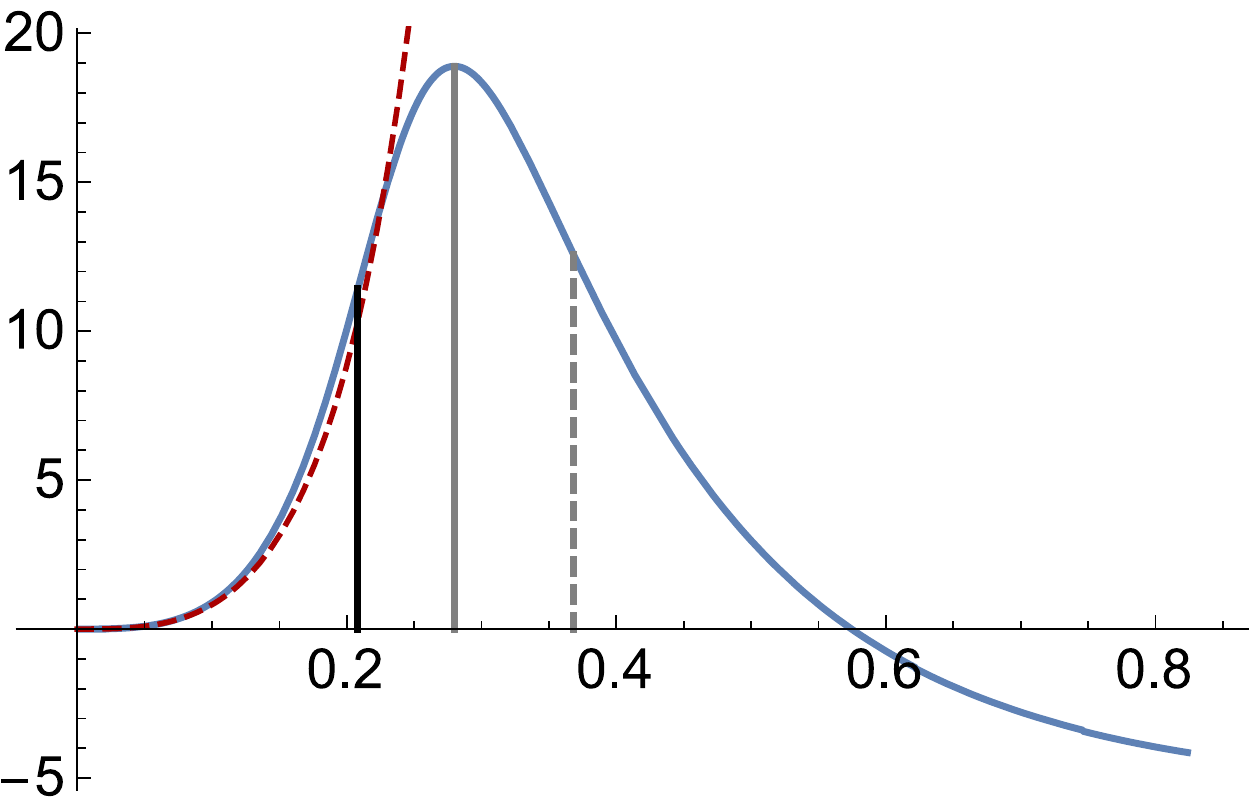} 
\put(-185,130){\large $\cal C_V$}
\put(5,20){\large $\cal T$}
\end{subfigure}\hspace{10mm}
\begin{subfigure}{.42\textwidth}
\includegraphics[width=\textwidth]{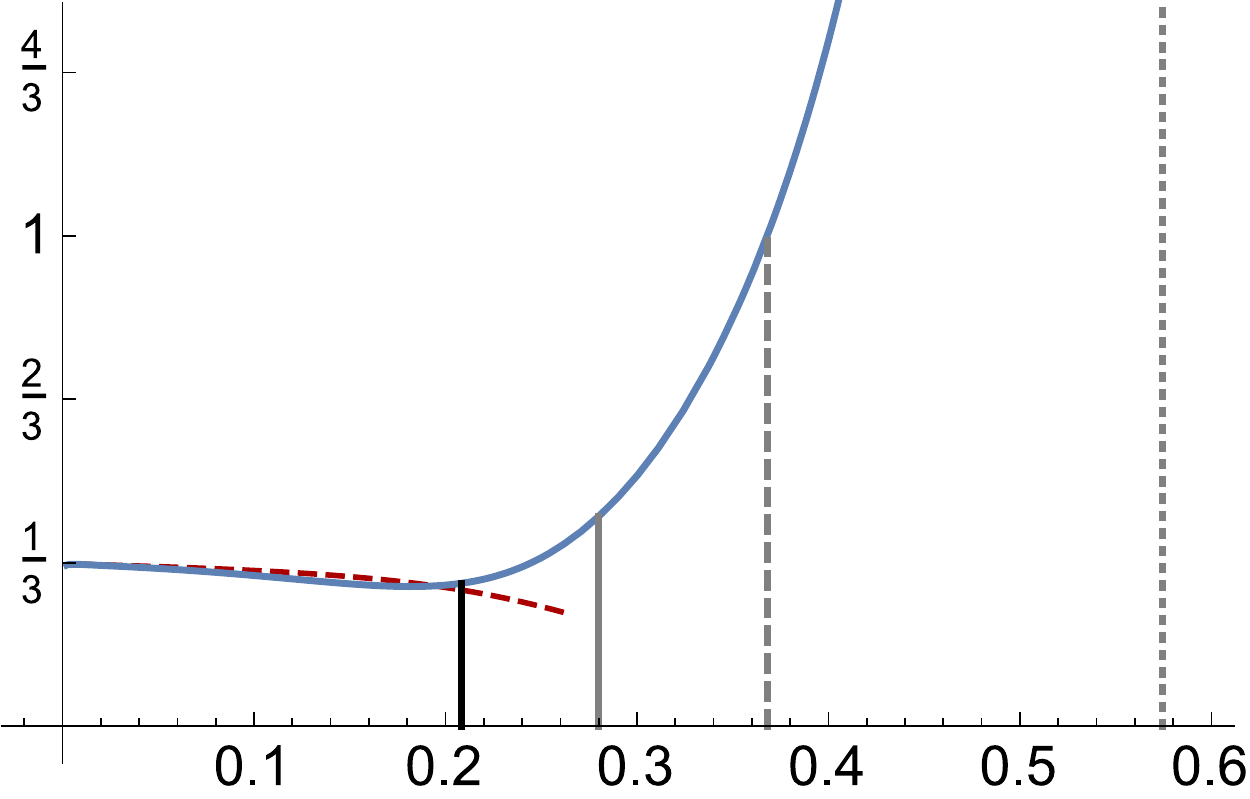} 
\put(-175,120){\large $v_s^2$}
\put(5,8){\large $\cal T$}
\end{subfigure}
\caption{Dimensionless specific heat and speed of sound as a function of the dimensionless temperature. The dashed, red curves correspond to the values derived from the low-temperature approximation \eq{eq.d3d7qgpresult}.}\label{fig.specheat}
\end{center}
\end{figure}

As is clear from Fig.~\ref{fig.thermo}, the transition from the low- to the high-temperature behavior takes place around $\mathcal{T}\sim 1$. Around this point several interesting features of the solution arise. We have marked the temperatures at which each of them takes place with a vertical line in Figs.~\ref{fig.thermo}, \ref{fig.dilvsT}, and \ref{fig.specheat}. In order of increasing temperature these features are as follows:
\begin{enumerate}
\item 
At a temperature ${\cal T}_1=0.21$, denoted with a black solid line in the figures, the  parameter controlling the backreaction of the flavor becomes 1, i.e. 
\be
\label{around}
\Qf \,e^{\phi_\mt{h}} = 1\,.
\ee
Note that at this temperature the horizon is well within the region where supergravity is valid since, in view of \eq{profile}, this region  is roughly speaking the analog of the $\vr \sim 1$ region in the zero-temperature solution. Supergravity is also valid at the three other regions that we will discuss below, since the corresponding temperatures  are parametrically of the same order. At ${\cal T}_1$ the speed of sound \eq{sound} is $v_s^2 ({\cal T}_1)= 0.29$. 

\item  
At ${\cal T}_2=0.28$, marked with a gray solid line in the figures, the specific heat \eq{heat} has a maximum, as shown in Fig.~\ref{fig.specheat}(left). At  temperatures below $\mathcal{T}_2$ the specific heat is a monotonically increasing function of the temperature, whereas it becomes monotonically decreasing at higher temperatures. It is interesting that at ${\cal T}_2$ the speed of sound exceeds the conformal value, since  
\be
v_s^2(\mathcal{T}_2) = 0.43>1/3 \,.
 \ee
 Note that the conformal value is recovered at  ${\cal T}=0$, as expected from \eq{eq.d3d7qgpresult}, which determines the low-temperature speed of sound to be
\be
v_s^2 = \frac{1}{3}  \left[ 1 - \frac{1}{6} \left( \Qf \,e^{\phi_\mt{h}({\cal T})} \right)^2 + \cdots \right] \,.
\ee

\item 
At  ${\cal T}_3=0.37$, marked with a dashed grey line in the figures, the speed of sound \eq{sound} becomes equal to the speed of light, as shown in Fig.~\ref{fig.specheat}(right). 
\item 
At ${\cal T}_4=0.57$, marked with a dotted grey line in the figures,  the specific heat becomes negative, as clearly seen in Fig.~\ref{fig.specheat}(left). Therefore at this point the system becomes locally thermodynamically unstable. From Eqs.~\eq{heat} and \eq{sound} we also see that at this temperature the entropy density attains a maximum as a function of the temperature, and the speed of sound diverges. 
\end{enumerate}

We will discuss these results further in the next section.

%%%%%%%%%%%%%%%
%%%%%%%%%%%%%%%%%%%%%%%%%%%%%%%%
\section{Discussion}
\label{disc}
%%%%%%%%%%%%%%%%%%%%%%%%%%%%%%%%
%%%%%%%%%%%%%%%

Our study of the gravity dual of a gauge theory with a Landau pole has  identified several features consistent with the presence of a maximum energy scale in the theory. For example, the energy \eq{mmax} of a string stretching along the entire range of the holographic coordinate is finite, and in fact may be used to define the scale of the Landau pole  itself. Similarly, the area in Planck units of slices of constant holographic coordinate attains a maximum at a scale slightly below the Landau pole, meaning that there is a maximum density of degrees of freedom per unit volume that the theory can describe. Ultimately, this property follows from the fact that the $g_{tt}$ component of the effective five-dimensional metric \eq{eq.LandaupoleasHV} exhibits non-monotonic behavior as a function of the holographic coordinate, as illustrated in \Fig{fig.5dmetricsusy}. In contrast,  the $G_{tt}$ component of the ten-dimensional metric, either in string or in Einstein frame, is  monotonic. In other words, the behavior of $g_{tt}$ arises from the behavior of the volume of the internal directions of the ten-dimensional geometry. 

Interestingly, this difference between the five- and the ten-dimensional metrics seems to be captured by two gauge-invariant, non-local observables in the gauge theory, the quark-antiquark potential and the EE. The former is sensitive to the ten-dimensional string-frame metric, since this is the metric that the hanging string couples to. Correspondingly, it does not exhibit any phase transitions as a function of the quark-antiquark separation. In contrast, the EE  is computed from an extremal surface in the five-dimensional metric, and it does exhibit 
multivaluedness and a phase transition as a function of the width of the boundary region, as can be seen in \Fig{fig:EEsusy}. 
%DMDM
This behavior is not purely a LP effect in the sense that the near-LP HV metric gives the single-valued and simple result \eq{LPEE}. Instead, the phase transition is a result of ``gluing'' together the IR metric in which $g_{xx}$ increases with the energy scale and a UV metric in which $g_{xx}$ decreases with the energy scale. In other words, the phase transition is a consequence of the non-monotonicity of the five-dimensional metric and it occurs around the scale at which $g_{xx}$ attains a maximum. Note also that, unlike in the case of confining backgrounds discussed in \cite{Klebanov:2007ws}, the phase transition that we have encountered takes place between two connected configurations of the holographic surface, as opposed to between a connected and a disconnected configuration.

Ref.~\cite{Jones:2016iwx} discussed the calculation of the EE using a higher-dimensional surface that entirely wraps the internal directions of the ten-dimensional geometry. In a full version of this calculation one would explicitly see that the multivaluedness of the EE is associated to the behavior of the internal components of the metric. However, the multivaluedness cannot be seen in the specific  calculation in \cite{Jones:2016iwx} because this was done perturbatively in the number of flavors. This means that it is valid for widths much larger than $1/\lp$, whereas the multivaluedness  appears at widths of order $1/\lp$.

The maximum number of degrees of freedom in the system can also be seen in the entropy density of thermal solutions, which attains a maximum as a function of the temperature, as illustrated in \Fig{fig.thermo}. This  property is intimately connected with an instability of the finite-temperature physics at high temperatures, namely  the fact that the specific heat becomes negative above the temperature 
${\cal T}_4$ defined in the previous section, as illustrated in \Fig{fig.specheat}. It is remarkable that this temperature lies in the regime where supergravity is applicable.  This thermal instability may be seen as a feature to be avoided from the viewpoint of phenomenological applications. In contrast, from the perspective of our motivation we view this situation as a holographic success, because it means that  interesting high-energy features of the gauge theory, including its possible thermal instabilities, \emph{can} be reliably studied via supergravity. 

Another property of the thermal solutions that holography can  predict is the fact that the speed of sound exceeds the conformal value at some temperature below the temperature ${\cal T}_2$ at which the specific heat attains its maximum value. Remarkably,  this happens before any thermal instability takes place, since 
${\cal T}_2 < {\cal T}_4$. It would be interesting to investigate possible phenomenological applications of this result, as well as possible connections with \cite{Hoyos:2016cob}.  In particular, a large value of the speed of sound codifies a stiff  equation of state, which could provide an explanation for the existence of neutron stars with a certain mass-size relationship \cite{Bedaque:2014sqa}.

At the qualitative level, all the features discussed above follow exclusively from the fact that the effective five-dimensional metric \eq{eq.LandaupoleasHV} near the Landau pole is a HV metric with HV exponent $\theta > p$, where $p$ is the number of spatial dimensions of the gauge theory. This suggests that many of these features may actually be rather universal among the class of Landau pole-afflicted gauge theories with a gravity dual. Moreover, as shown in Appendix \ref{continuation}, holographic renormalization can be fully implemented for all such theories by analytically continuing the gravitational solutions to an AdS solution. 
%DMDM
We emphasize that this procedure does not provide a UV completion of the theory but simply a prescription to renormalize a UV-incomplete theory. 

Although we have focused on the case of massless quarks throughout the paper, it is easy to understand how the physics would be modified in the presence of a non-zero quark mass $M_q$. This modification only depends on the ratio 
$M_q /\lp$ and, at a qualitative level, amounts to cutting off the geometry at the scale $M_q$ at which the quarks decouple,  and replacing the region  below  by pure AdS. Thus, if $M_q \lesssim \lp$ then one only sees the Landau pole geometry glued to pure AdS in the IR. In contrast, if $M_q \ll \lp$ then one sees essentially the same solution that we have discussed except in the deep IR, where the log-AdS geometry eventually turns into pure AdS. In all cases, as long as $M_q$ is non-zero, the singularity in the IR gets replaced by the smooth AdS geometry. 

We close with a possibility suggested by Quantum Electrodynamics (QED). Since the one-loop $\beta$-function of QED is positive, at the perturbative level one concludes that the theory possesses a Landau pole. However, recent non-perturbative studies \cite{Gockeler, Kim,Gies} suggest a more elaborate picture. The key non-perturbative insight is that, above a certain value of the bare   coupling, spontaneous chiral symmetry breaking takes place. As a consequence the renormalized electron mass is non-zero even if the bare mass is set to zero. This results in a region in the renormalized mass--coupling plane that is physically excluded in the sense that it cannot be obtained from any value of the bare parameters. The putative Landau pole would lie in the excluded region. However, QED is still trivial even at the non-perturbative level because the only point in the allowed region where the continuum limit can be taken corresponds to a vanishing value of the renormalized coupling. The key point is that the strong-coupling physics triggers the spontaneous breaking of a symmetry which modifies the naive perturbative expectation. It would be interesting to investigate whether a similar picture could be realized in a theory with a gravity dual, in particular whether some sort of symmetry breaking (chiral or otherwise) could change the UV behavior of the solution and render it regular. This could be investigated in our current set-up by considering a more general, less symmetric  ansatz than the one that we have studied.

%%%%%%%%%%%%%%%
%%%%%%%%%%%%%%%%%%%%%%%%%%%%%%%%
\section*{Acknowledgements}
%%%%%%%%%%%%%%%%%%%%%%%%%%%%%%%%
%%%%%%%%%%%%%%%

%DMDM
We thank Bartomeu Fiol, Jaume Garriga,  Daniele Musso, and specially Carlos Nu\~nez and Ioannis Papadimitriou,  for discussions and comments. We are supported by grants 2014-SGR-1474, MEC FPA2013-46570-C2-1-P, MEC FPA2013- 46570-C2-2-P, CPAN CSD2007-00042 Consolider-Ingenio 2010, ERC Starting Grant HoloLHC-306605 and Maria de Maeztu Unit of Research Excellence distinction. JT is supported by the  Advanced ARC project ``Holography, Gauge Theories and Quantum Gravity'' and by the Belgian Fonds National de la Recherche Scientifique FNRS (convention IISN 4.4503.15). 

\appendix
\section{Holographic renormalization}
\label{continuation}
Let us consider a simple model of gravity plus a scalar in $D$ dimensions
\begin{equation}\label{gravscalar}
S\,=\,\frac{1}{2\kappa_D^2}\,\int\,\left(R*1-\frac12\dd\Phi\wedge*\dd\Phi-V\left(\Phi\right)*1\right)\,,
\end{equation}
with an exponential type potential 
\begin{equation}
V\left(\Phi\right)\,=\,V_0 \,e^{\alpha \Phi}\,,
\end{equation}
for some constants $V_0$ and $\alpha$. This action has a hyperscaling violating solution
\begin{eqnarray}
\dd s_D^2&=&\bigg(\frac{r}{\mathsf{R}}\bigg)^{-\frac{2\theta}{p}}\left[\bigg(\frac{r}{\mathsf{R}}\bigg)^2\dd x_{1,p}^2+\left(\frac{\mathsf{R}}{r}\right)^2\dd r^2\right]\,,\nonumber\\[2mm]
e^{\alpha\Phi}&=&\bigg(\frac{r}{\mathsf{R}}\bigg)^{\frac{2\theta}{p}}\,,
\end{eqnarray}
where clearly $D=p+2$ and the parameters in the solution are related to those in the action through
\begin{equation}
\label{ttt}
\theta\,=\,\frac{p^2\,\alpha^2}{p\,\alpha^2-2}\,,\qquad\qquad\qquad \mathsf{R}^2\,=\,-\frac{2p\left[2+p\left(2-\alpha^2\right)\right]}{V_0\left(p\,\alpha^2-2\right)^2}\,.
\end{equation}
%DMDM
Suppose now that $q$ out of the $p$ spatial directions are compact of size $l$ and let us reduce on this $q$-torus. The ansatz for the reduced metric to be in Einstein frame is thus
\begin{equation}
\dd s_D^2\,=\,e^{\frac{2d}{(q-p)}\varphi}\,\dd s_{D-q}^2+e^{2\varphi}\,\dd x_q^2\,.
\end{equation} 
Defining $\kappa_{D-q}^2=\left(2\pi l\right)^q\kappa_D^2$ we get the reduced action
\begin{equation}\label{reduced}
S\,=\,\frac{1}{2\kappa_{D-q}^2}\,\int\,\left(R*1-\frac12\dd\Phi\wedge*\dd\Phi-\frac{p\,q}{(p-q)}\dd\varphi\wedge*\dd\varphi-V\left(\Phi\right)\,e^{\frac{2q}{(q-p)}\varphi}\,*1\right)\,.
\end{equation}
It is consistent to truncate this action to a single scalar by identifying $-\alpha p\varphi=\Phi$ and, after redefining 
\begin{equation}
\left[1+\frac{2q}{\alpha^2p\left(p-q\right)}\right]\Phi^2\,=\,\tilde{\Phi}^2\,,
\end{equation}
we get a ($D-q$)-dimensional action of the form (\ref{gravscalar}) with a modified parameter
\begin{equation}
\tilde{\alpha}\,=\,\alpha\left[1+\frac{2q}{\alpha^2p\left(p-q\right)}\right]^{1/2}\,.
\end{equation}
It is then immediate that this action admits a hyperscaling violating solution with $p-q$ spatial directions and coefficient
\begin{equation}
\tilde{\theta}\,=\,\frac{\left(p-q\right)^2\tilde{\alpha}^2}{\left(p-q\right)\tilde{\alpha}^2-2}\,=\,\theta-q\,.
\end{equation}
This hyperscaling solution reduces to AdS if $q=\theta$. Notice the curious fact that the AdS solution exists even if the potential in the reduced action (\ref{reduced}) does not seem to admit any critical points. This happens because for $\theta=q$ the truncation gives a constant potential and the system reduces to the Einstein--Hilbert term plus a cosmological constant. For our particular value of $\theta=7/2$ this  AdS space would have fractional dimension 
\be
D-\theta \,=\, 5- \frac{7}{2} \,=\, \frac{3}{2} \,, 
\label{deff}
\ee
which would be dual to a gauge theory in $d_\mt{eff}=1/2$ dimensions, as in \eqq{deffmain}.

The reduced action can be renormalized by including the standard counterterm
\begin{equation}
S_{\rm ct}\,=\,-\frac{1}{2\kappa_{D-q}^2}\,\int\,\sqrt{-\gamma_{D-q-1}}\,\,\widetilde{W}\,,
\end{equation}
with $\gamma_{D-q-1}$ the induced metric on the $\left(D-q-1\right)$-dimensional boundary. The superpotential $\widetilde{W}$ is related to the potential as
\begin{equation}
\widetilde V\,=\,\frac{1}{2} \left(\frac{\dd \widetilde{W}}{\dd \tilde{\Phi}}\right)^2-
\frac{1}{4} \left( \frac{D-q-1}{D-q-2} \right)\,\widetilde{W}^2\,
\end{equation}
and reads explicitly 
\begin{equation}
\widetilde{W}=2 \left(-\frac{2V_0\left(D-q-2\right)}{2-\left(D-q-2\right)\left(\tilde{\alpha}^2-2\right)}\right)^{1/2}\,e^{\frac{\tilde{\alpha}}{2}\tilde{\Phi}}\,.
\end{equation}
This counterterm can be uplifted on the $q$-torus and gives precisely
\begin{equation}
S_{\rm ct}\,=\,-\frac{1}{2 \kappa_{D}^2}\,\int\,\sqrt{-\gamma_{D-1}}\,\,W\,,
\end{equation}
with
\begin{equation}
W\,=2\left(-\frac{2V_0\left(D-2\right)}{2-\left(D-2\right)\left(\alpha^2-2\right)}\right)^{1/2}\,e^{\frac{\alpha}{2}\Phi}\,,
\end{equation}
that is, the superpotential associated with the original action (\ref{gravscalar}), which of course renormalizes it.

These considerations can be extended to the case of several scalars, when the hyperscaling solution needs not be exact but only asymptotic. One can see that the leading exponential, which can be made of a combination of scalars, is the only one that matters, and all the considerations above apply to it. 

In the particular case under study in this paper it is easy to check that the two leading terms in our potential \eq{potpot} arise from two contributions of similar size that can be traced back to the kinetic terms of the $F_5$ and $F_1$ RR forms  in the ten-dimensional supergravity  action. Specifically we have
\begin{equation}
V \simeq \frac{ 1 }{ 8\cdot 2^{2/3} L^2 } \left( e^{- \frac{8}{3} \psi} + \Qf^2 e^{2\phi - \frac{8}{5}\chi - \frac{16}{15} \psi } \right) \simeq  \frac{ 1 }{ 4\cdot 2^{2/3} L^2 }  e^{- \frac{8}{3} \psi} \equiv V_0 \,e^{\alpha \Phi} \,,
\end{equation}
which can be obtained from the superpotential
\begin{equation}
W \simeq \frac{1}{2^{4/3} L} \left( e^{-\frac{4}{3} \psi} + \Qf e^{\phi - \frac{4}{5} \chi - \frac{8}{15} \psi} \right) \simeq \frac{1}{2^{1/3} L} \ e^{-\frac{4}{3} \psi} = 2 \sqrt{V_0} \, e^{\frac{\alpha}{2} \Phi}\,.
\end{equation}
To read the correct value of $\alpha$ we need to properly normalize the kinetic term for the scalars, as in (\ref{gravscalar}). At leading order in the UV we get the relation $\psi=\pm\left(21/32\right)^{1/2}\Phi$, which gives the parameter $\alpha=\mp\left(14/3\right)^{1/2}$ required to reproduce the hyperscaling coefficient $\theta=7/2$.

\section{Calculation of the quark-antiquark potential}
\label{wilson}
A natural observable to study is the Wilson loop. We perform a simple calculation for a time invariant configuration of two external sources placed on the boundary, separated by a distance $L$, and a string hanging between them in the bulk.
We choose a parametrisation on the world-sheet by coordinates $\sigma$ and $\tau$. The string world-sheet in the bulk is then given by an embedding $X^M(\sigma, \tau)$ and the action is
\begin{equation}
S_\mt{NG}=\frac{1}{2 \pi \ell_s^2}\int_Cd\sigma d\tau\sqrt{\det_{ab}(G^\mt{st}_{MN}\partial_a X^M\partial_bX^N})\,.
\end{equation}
We choose $t=\tau, x=\sigma, \varrho=R(\sigma)\equiv R(x)$ and we obtain the following action
\begin{equation}
S_\mt{NG}=\frac{1}{2\,\pi\,l_s^2}\int_{-L/2}^{L/2}\int_0^Tdt dx\sqrt{G^\mt{st}_{tt}G^\mt{st}_{xx}+G^\mt{st}_{\varrho\varrho}G^\mt{st}_{tt} R'^2}\,.
\end{equation}
Since the problem is  invariant under translations in $x$, the ``energy''
\begin{equation}
H=\frac{\delta \mathcal{L}}{\delta(\partial_x R)}\partial_x R-\mathcal{L}=\frac{G^\mt{st}_{xx}\sqrt{G^\mt{st}_{tt}}}{\sqrt{G^\mt{st}_{xx}+G^\mt{st}_{\varrho\varrho} R'^2}}\,,
\end{equation} 
with $\mathcal{L}$ is the Nambu-Goto Lagrangian, is conserved. Its value can be computed at the turning point of the string, which for convenience is taken to be at $x=0$ and 
$\varrho=\varrho_0$. Since at that point $R'(0)=0$, we have $H=\sqrt{G^\mt{st}_{xx}(0) G^\mt{st}_{tt}(0)}$. This leads us to the following differential equation for $R$
\begin{equation}
R'=\sqrt{\frac{G^\mt{st}_{xx}}{G^\mt{st}_{\varrho\varrho}}}\frac{\sqrt{G^\mt{st}_{xx}G^\mt{st}_{tt}-G^\mt{st}_{xx}(0) G^\mt{st}_{tt}(0)}}{\sqrt{G^\mt{st}_{xx}(0) G^\mt{st}_{tt}(0)}}\,.
\end{equation}
Using the UV expansion \eqref{eq.UVnumerics} one can show that close to the Landau-Pole  $R'(x)\to \infty$.  The significance of this is that the string ends at the boundary perpendicularly, despite the boundary being located at finite proper distance, and thus there is no extremal force being exerted on the quarks. The length of the string stretching between the two quarks is given by
\begin{equation}
\frac{L}{2}=\int_0^{L/2} dx=\int_{\varrho_0}^{\infty} \frac{d\varrho}{R'}\, ,
\end{equation}
while the energy of the system is evaluated by calculating the on-shell action divided by the integral over time as follows
\begin{equation}
\label{eee}
 E=\frac{S_\mt{NG}}{\Delta t}=2\,\int_{\varrho_0}^\infty \sqrt{ G^\mt{st}_{tt}(G^\mt{st}_{xx}+G^\mt{st}_{\varrho\varrho} R'^2)}
 \, \frac{d \varrho}{ R'}\,.
\end{equation}
In contrast with the usual Wilson loop calculation, the above quantity is in fact finite and the integrand goes to zero in the UV of our geometry. This is consistent with the existence of a UV cut-off set by the Landau-Pole scale, in the sense that the quarks can not be infinitely massive. 

As was explained around equation \eqref{eq.scalings}, the dependence of the supergravity equations of motion on the charges of the theory, $\Qc,\Qf$, can be removed through appropriate rescalings. In particular, we perform the following rescalings of the metric functions
\begin{align}\label{eq.scalingsmetric}
&G^\mt{st}_{tt}=\Qf^{-1/2} \bar G^\mt{st}_{tt}\,,\quad G^\mt{st}_{xx}=\Qf^{-1/2} \bar G^\mt{st}_{xx}\,,\quad 
G^{st}_{\rho\rho}=\Qf^{-1/2} \Qc^{1/2} \bar G^{st}_{\rho\rho} \,.
\end{align}
In order to reduce the Nambu-Goto equations of motion to the same dependence we must further rescale the $x$-coordinate as  $x\to \Qc^{1/4}\, x$. Under these circumstances the Nambu-Goto action scales homogeneously as 
\be
S_\mt{NG}=\Qf^{-1/2}\,\Qc^{1/4} \bar S_\mt{NG} \,,
\ee
where $\bar S_\mt{NG}$ is given by the same expression as $S_\mt{NG}$ but with $\Qc, \Qf$ set to unity. One immediate consequence of this is that the distance between the endpoints of the strings satisfies the scaling property
\begin{equation}
\label{redefL}
L=\Qc^{1/4}\, \bar L\,,
\end{equation}
where $\bar L$ is given by the same expression as $L$ but with $\Qc, \Qf$ set to unity. 
The energy of the system, which is evaluated by calculating the on-shell action divided by the integral over time, can also be expressed in terms of dimensionless variables as
\begin{equation}
\label{redefE}
E=\frac{S_\mt{NG}}{\Delta t}=\frac{\Qc^{1/4}}{\Qf^{1/2}}\frac{\bar S_\mt{NG}}{\Delta t}\equiv \frac{1}{2\pi \ell_s^2} \frac{\Qc^{1/4}}{\Qf^{1/2}} \, \bar E\,,
\end{equation}
where
\begin{equation}
\bar E=2\,\int_{\varrho_0}^\infty \sqrt{\bar G^\mt{st}_{tt}(\bar G^\mt{st}_{xx}+\bar G^\mt{st}_{\varrho\varrho} \bar R'^2)} \, \frac{d\varrho}{\bar R'}\,.
\end{equation}
Applying the above procedure for the Landau-Pole geometry, we obtain
\begin{equation}
\bar L\sim \varrho_0^{2}\,, \quad \bar E\sim \varrho_0\,.
\end{equation}

\section{Calculation of the entanglement entropy}
\label{ent}
\begin{figure}[t]
\begin{center}
\begin{subfigure}{.45\textwidth}
\includegraphics[width=\textwidth]{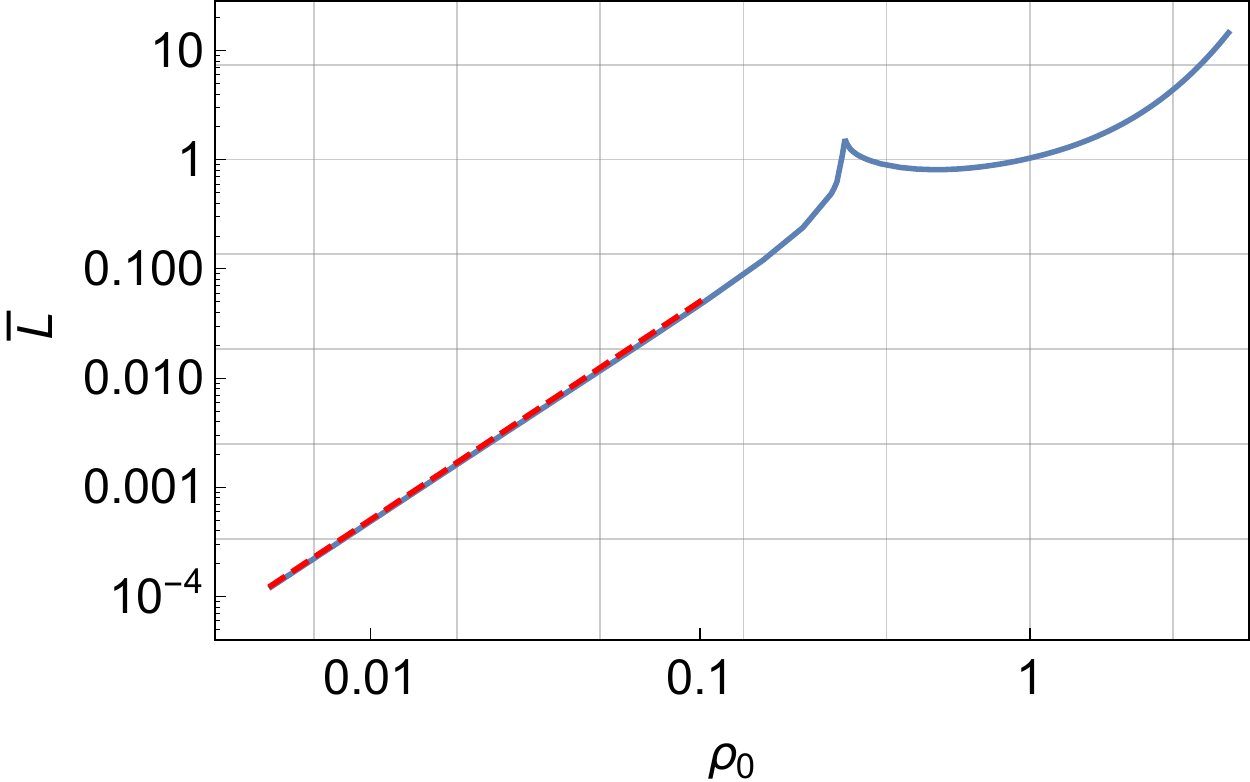} 
\end{subfigure}\hspace{5mm}
\begin{subfigure}{.45\textwidth}
\includegraphics[width=\textwidth]{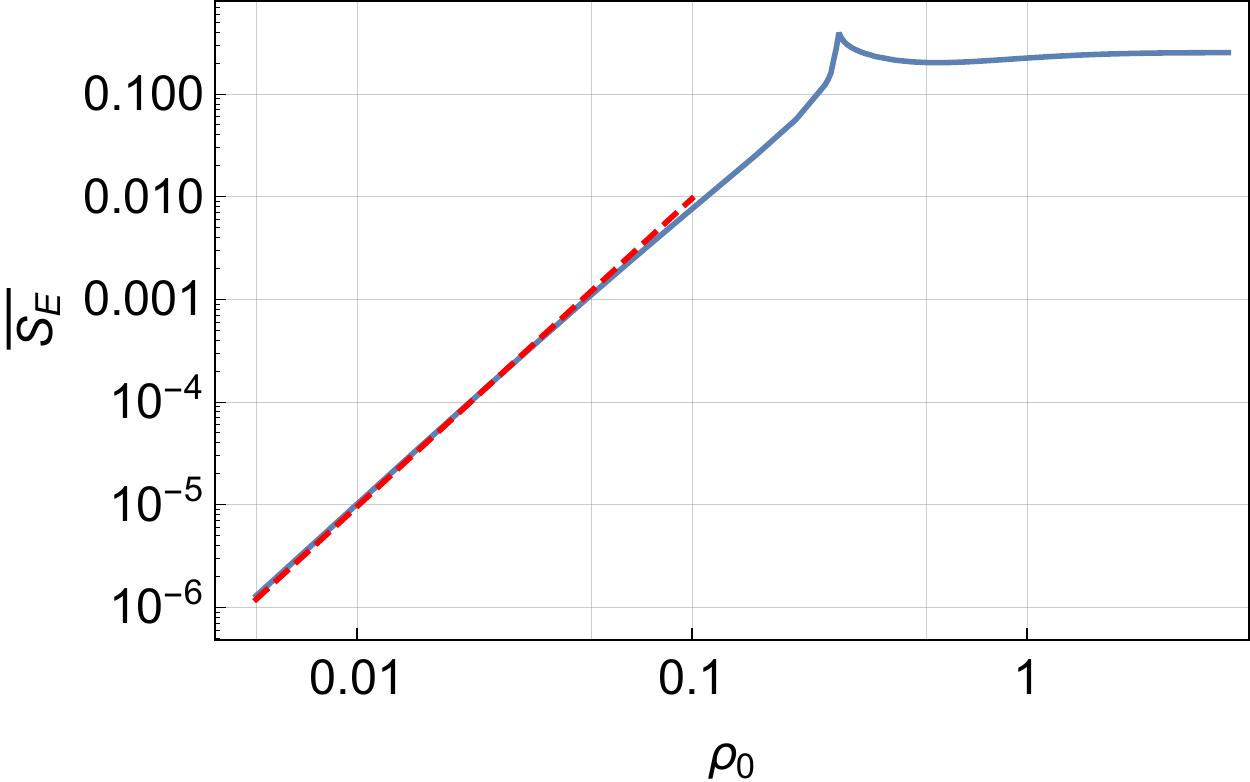} 
\end{subfigure}\hspace{10mm}
 \caption{Width (left) and EE (right) as a function of the penetration depth $\varrho_0$ of the extremal surface in the bulk.}
 \label{internal}
\end{center}
\end{figure}

To define the entanglement entropy of the boundary field theory, we start by dividing a boundary constant-time, spatial slice into two regions, A and B. In what follows we will choose region A to be an infinitely long stripe of width $L$, described by $x_1\, \in \, [-L/2, +L/2]$, $x_2 \,\in\, (-\infty, +\infty)$ and $x_3 \,\in\, (-\infty, +\infty)$, and region B to be its complement. According to the holographic dictionary, the entanglement entropy is then given by the area of the minimal surface in the bulk that is anchored on the boundary of A, $\partial$A. 
We choose a parametrisation of the minimal surface by coordinates $\sigma_1$, $\sigma_2$ and $\sigma_3$. The minimal surface  in the bulk is then given by an embedding $X^M(\sigma_1, \sigma_2, \sigma_3)$ and the action is
\begin{equation}
S=\frac{1}{2 \pi \ell_s^2}\int_Cd\sigma d\tau\sqrt{\det_{ab}(g_{MN}\partial_a X^M\partial_bX^N})\,,
\end{equation}
where $g$ is the effective five-dimensional metric \eq{eq.5dmetricdef}.
We choose 
\be
x_1=\sigma_1 \sac x_2=\sigma_2 \sac x_3=\sigma_3 \sac 
\varrho=R(\sigma_1)\equiv R(x_1)
\ee
 and we obtain the following action
\begin{equation}
S=\frac{L_2 L_3 }{2 \pi \ell_s^2}\,\int_{-L/2}^{L/2}\, dx_1\, g_{xx}\sqrt{g_{xx}+g_{\varrho\varrho} R'^2}\,,
\end{equation}
where $L_i=\int_{-\infty}^{\infty} dx_i$. Since the problem is  invariant under translations in $x$, the ``energy''
\begin{equation}
H=\frac{\delta \mathcal{L}}{\delta(\partial_x R)}\partial_x R-\mathcal{L}=\frac{g_{xx}\,^2}{\sqrt{g_{xx}+g_{\varrho\varrho} R'^2}}
\end{equation} 
is conserved. Its value can be computed at the turning point of the string, which for convenience is taken to be at $x_1=0$ and $\varrho=\varrho_0$. Since at that point $R'(0)=0$, we have $H=g_{xx}(0)^{3/2}$. This leads us to the following differential equation for $R$
\begin{equation}
R'=\sqrt{\frac{g_{xx}}{g_{\varrho\varrho}}}\frac{\sqrt{g_{xx}\,^3-g_{xx}(0)^3}}{\sqrt{g_{xx}(0)^3}}\,
\end{equation}
for which we get
\begin{equation}
L=2\,\int_0^{L/2} dx_1=2\,\int_{\varrho_0}^{\infty} \frac{d\varrho}{R'}\,\equiv Q_c^{1/4}\,\bar L\,.
\end{equation}
The entanglement entropy density associated with region A is evaluated by calculating the on-shell action
\begin{equation}
\label{redefS}
S_{E}=\frac{1}{4 G}\frac{S}{L_2 L_3}=\frac{1}{4 G}\frac{2}{2\pi \ell_s^2}
\int_{\varrho_0}^\infty g_{xx}\,\sqrt{g_{xx}+g_{\varrho\varrho} R'^2}\,\frac{d\varrho}{R'}\,\equiv \frac{1}{4 G}\frac{1}{2\pi\ell_s^2} \frac{\Qc^{1/4}}{\Qf^{3/4}}\,\bar S_E.
\end{equation}
In the above 
\begin{align}
&\bar L=2\,\int_{\varrho_0}^{\infty} \frac{d \varrho}{\bar R'}\,,\nonumber\\[2mm]
&\bar S_E=2\int_{\varrho_0}^\infty \bar g_{xx}\,\sqrt{\bar g_{xx}+\bar g_{\varrho\varrho} \bar R'^2}\,\frac{d\varrho}{\bar R'}
\end{align}
 are dimensionless, and to obtain them we have used the scaling \eqref{eq.scalingsmetric}. Applying the above procedure to the Landau-Pole geometry we find
\begin{equation}
\bar L\sim \varrho_0^{2}\,,\quad \bar S_E\sim \varrho_0^{3}\,,
\end{equation}
which is valid for small $\varrho_0$. The result for arbitrary $\varrho_0$ is shown in \Fig{internal}.
We see that in a certain range of values of $\bar L$ there is more than one possible value of $\varrho_0$, i.e.~there is more than one extremal surface. This results in the EE being a multivalued function of $L$, as shown in \Fig{fig:EEsusy}.


\begin{thebibliography}{99} 

\bibitem{Karch:2002sh} 
  A.~Karch and E.~Katz,
  ``Adding flavor to AdS / CFT,''
  JHEP {\bf 0206}, 043 (2002)
  doi:10.1088/1126-6708/2002/06/043
  [hep-th/0205236].

\bibitem{Kruczenski:2003be} 
  M.~Kruczenski, D.~Mateos, R.~C.~Myers and D.~J.~Winters,
  ``Meson spectroscopy in AdS / CFT with flavor,''
  JHEP {\bf 0307}, 049 (2003)
  doi:10.1088/1126-6708/2003/07/049
  [hep-th/0304032].
  
  
  \bibitem{Babington:2003vm} 
  J.~Babington, J.~Erdmenger, N.~J.~Evans, Z.~Guralnik and I.~Kirsch,
  ``Chiral symmetry breaking and pions in nonsupersymmetric gauge / gravity duals,''
  Phys.\ Rev.\ D {\bf 69}, 066007 (2004)
  doi:10.1103/PhysRevD.69.066007
  [hep-th/0306018].
  
  \bibitem{Erdmenger:2007cm} 
  J.~Erdmenger, N.~Evans, I.~Kirsch and E.~Threlfall,
  ``Mesons in Gauge/Gravity Duals - A Review,''
  Eur.\ Phys.\ J.\ A {\bf 35}, 81 (2008)
  doi:10.1140/epja/i2007-10540-1
  [arXiv:0711.4467 [hep-th]].
  
  \bibitem{CasalderreySolana:2011us} 
  J.~Casalderrey-Solana, H.~Liu, D.~Mateos, K.~Rajagopal and U.~A.~Wiedemann,
  ``Gauge/String Duality, Hot QCD and Heavy Ion Collisions,''
  arXiv:1101.0618 [hep-th].
  
  
  
  %\cite{Benini:2006hh}
\bibitem{Benini:2006hh}
  F.~Benini, F.~Canoura, S.~Cremonesi, C.~Nunez and A.~V.~Ramallo,
  ``Unquenched flavors in the Klebanov-Witten model,''
  JHEP {\bf 0702} (2007) 090
  [hep-th/0612118].
  %%CITATION = HEP-TH/0612118;%%



  
    %\cite{Nunez:2010sf}
\bibitem{Nunez:2010sf}
  C.~Nunez, A.~Paredes and A.~V.~Ramallo,
  ``Unquenched Flavor in the Gauge/Gravity Correspondence,''
  Adv.\ High Energy Phys.\  {\bf 2010} (2010) 196714
  doi:10.1155/2010/196714
  [arXiv:1002.1088 [hep-th]].
  %%CITATION = doi:10.1155/2010/196714;%%


  
  %\cite{Bigazzi:2005md}
\bibitem{Bigazzi:2005md}
  F.~Bigazzi, R.~Casero, A.~L.~Cotrone, E.~Kiritsis and A.~Paredes,
  ``Non-critical holography and four-dimensional CFT's with fundamentals,''
  JHEP {\bf 0510} (2005) 012
  doi:10.1088/1126-6708/2005/10/012
  [hep-th/0505140].
  %%CITATION = doi:10.1088/1126-6708/2005/10/012;%%



  %\cite{Bigazzi:2009bk}
\bibitem{Bigazzi:2009bk}
  F.~Bigazzi, A.~L.~Cotrone, J.~Mas, A.~Paredes, A.~V.~Ramallo and J.~Tarrio,
  ``D3-D7 Quark-Gluon Plasmas,''
  JHEP {\bf 0911} (2009) 117
  [arXiv:0909.2865 [hep-th]].
  %%CITATION = ARXIV:0909.2865;%%
  
  \bibitem{Kanitscheider:2008kd} 
  I.~Kanitscheider, K.~Skenderis and M.~Taylor,
  ``Precision holography for non-conformal branes,''
  JHEP {\bf 0809}, 094 (2008)
  doi:10.1088/1126-6708/2008/09/094
  [arXiv:0807.3324 [hep-th]].

\bibitem{Wiseman:2008qa} 
  T.~Wiseman and B.~Withers,
  ``Holographic renormalization for coincident Dp-branes,''
  JHEP {\bf 0810}, 037 (2008)
  doi:10.1088/1126-6708/2008/10/037
  [arXiv:0807.0755 [hep-th]].

\bibitem{Papadimitriou:2011qb} 
  I.~Papadimitriou,
  ``Holographic Renormalization of general dilaton-axion gravity,''
  JHEP {\bf 1108}, 119 (2011)
  doi:10.1007/JHEP08(2011)119
  [arXiv:1106.4826 [hep-th]].


\bibitem{Chemissany:2014xsa} 
  W.~Chemissany and I.~Papadimitriou,
  ``Lifshitz holography: The whole shebang,''
  JHEP {\bf 1501}, 052 (2015)
  doi:10.1007/JHEP01(2015)052
  [arXiv:1408.0795 [hep-th]].
  
  \bibitem{Aharony:2005zr} 
  O.~Aharony, A.~Buchel and A.~Yarom,
  ``Holographic renormalization of cascading gauge theories,''
  Phys.\ Rev.\ D {\bf 72}, 066003 (2005)
  doi:10.1103/PhysRevD.72.066003
  [hep-th/0506002].
  
  \bibitem{Bertolini:2015hua} 
  M.~Bertolini, D.~Musso, I.~Papadimitriou and H.~Raj,
  ``A goldstino at the bottom of the cascade,''
  JHEP {\bf 1511}, 184 (2015)
  doi:10.1007/JHEP11(2015)184
  [arXiv:1509.03594 [hep-th]].
  
    \bibitem{Kanitscheider:2009as} 
  I.~Kanitscheider and K.~Skenderis,
  ``Universal hydrodynamics of non-conformal branes,''
  JHEP {\bf 0904}, 062 (2009)
  doi:10.1088/1126-6708/2009/04/062
  [arXiv:0901.1487 [hep-th]].

%\cite{Gouteraux:2011ce}
\bibitem{Gouteraux:2011ce}
  B.~Gouteraux and E.~Kiritsis,
  ``Generalized Holographic Quantum Criticality at Finite Density,''
  JHEP {\bf 1112} (2011) 036
  doi:10.1007/JHEP12(2011)036
  [arXiv:1107.2116 [hep-th]].
  %%CITATION = doi:10.1007/JHEP12(2011)036;%%
  
  \bibitem{Faedo:2014ana} 
  A.~F.~Faedo, A.~Kundu, D.~Mateos and J.~Tarrio,
  ``(Super)Yang-Mills at Finite Heavy-Quark Density,''
  JHEP {\bf 1502}, 010 (2015)
  doi:10.1007/JHEP02(2015)010
  [arXiv:1410.4466 [hep-th]].

\bibitem{Faedo:2015ula} 
  A.~F.~Faedo, D.~Mateos and J.~Tarrio,
  ``Three-dimensional super Yang-Mills with unquenched flavor,''
  JHEP {\bf 1507}, 056 (2015)
  doi:10.1007/JHEP07(2015)056
  [arXiv:1505.00210 [hep-th]].
  
  
  \bibitem{Faedo:2015urf} 
  A.~F.~Faedo, A.~Kundu, D.~Mateos, C.~Pantelidou and J.~Tarrio,
  ``Three-dimensional super Yang-Mills with compressible quark matter,''
  JHEP {\bf 1603}, 154 (2016)
  doi:10.1007/JHEP03(2016)154
  [arXiv:1511.05484 [hep-th]].

%\cite{Faedo:2016jbd}
\bibitem{Faedo:2016jbd}
  A.~F.~Faedo, D.~Mateos, C.~Pantelidou and J.~Tarrio,
  %``Unquenched flavor on the Higgs branch,''
  JHEP {\bf 1611} (2016) 021
  doi:10.1007/JHEP11(2016)021
  [arXiv:1607.07773 [hep-th]].
  %%CITATION = doi:10.1007/JHEP11(2016)021;%%

\bibitem{Mateos:2011tv} 
  D.~Mateos and D.~Trancanelli,
  ``Thermodynamics and Instabilities of a Strongly Coupled Anisotropic Plasma,''
  JHEP {\bf 1107}, 054 (2011)
  doi:10.1007/JHEP07(2011)054
  [arXiv:1106.1637 [hep-th]].


  
  \bibitem{Klebanov:2000nc} 
  I.~R.~Klebanov and A.~A.~Tseytlin,
  ``Gravity duals of supersymmetric SU(N) x SU(N+M) gauge theories,''
  Nucl.\ Phys.\ B {\bf 578}, 123 (2000)
  doi:10.1016/S0550-3213(00)00206-6
  [hep-th/0002159].
  
  \bibitem{Benvenuti:2005qb} 
  S.~Benvenuti, M.~Mahato, L.~A.~Pando Zayas and Y.~Tachikawa,
  ``The Gauge/gravity theory of blown up four cycles,''
  hep-th/0512061.
  
  %\cite{Cotrone:2012um}
\bibitem{Cotrone:2012um}
  A.~L.~Cotrone and J.~Tarrio,
  ``Consistent reduction of charged D3-D7 systems,''
  JHEP {\bf 1210} (2012) 164
  [arXiv:1207.6703 [hep-th]].
  %%CITATION = ARXIV:1207.6703;%%

  

  %\cite{Benini:2007kg}
\bibitem{Benini:2007kg}
  F.~Benini,
  ``A Chiral cascade via backreacting D7-branes with flux,''
  JHEP {\bf 0810} (2008) 051
  doi:10.1088/1126-6708/2008/10/051
  [arXiv:0710.0374 [hep-th]].
  %%CITATION = doi:10.1088/1126-6708/2008/10/051;%%
  
  
  
%\cite{Aharony:1998xz}
\bibitem{Aharony:1998xz}
  O.~Aharony, A.~Fayyazuddin and J.~M.~Maldacena,
  ``The Large N limit of N=2, N=1 field theories from three-branes in F theory,''
  JHEP {\bf 9807} (1998) 013
  doi:10.1088/1126-6708/1998/07/013
  [hep-th/9806159].
  %%CITATION = doi:10.1088/1126-6708/1998/07/013;%%


  
    %\cite{Ogawa:2011bz}
\bibitem{Ogawa:2011bz}
  N.~Ogawa, T.~Takayanagi and T.~Ugajin,
  ``Holographic Fermi Surfaces and Entanglement Entropy,''
  JHEP {\bf 1201} (2012) 125
  doi:10.1007/JHEP01(2012)125
  [arXiv:1111.1023 [hep-th]].
  %%CITATION = doi:10.1007/JHEP01(2012)125;%%

  %\cite{Huijse:2011ef}
\bibitem{Huijse:2011ef}
  L.~Huijse, S.~Sachdev and B.~Swingle,
  ``Hidden Fermi surfaces in compressible states of gauge-gravity duality,''
  Phys.\ Rev.\ B {\bf 85} (2012) 035121
  doi:10.1103/PhysRevB.85.035121
  [arXiv:1112.0573 [cond-mat.str-el]].
  %%CITATION = doi:10.1103/PhysRevB.85.035121;%%
  
  
  
  %\cite{Dong:2012se}
\bibitem{Dong:2012se}
  X.~Dong, S.~Harrison, S.~Kachru, G.~Torroba and H.~Wang,
  ``Aspects of holography for theories with hyperscaling violation,''
  JHEP {\bf 1206} (2012) 041
  doi:10.1007/JHEP06(2012)041
  [arXiv:1201.1905 [hep-th]].
  %%CITATION = doi:10.1007/JHEP06(2012)041;%%
  
  
  %\cite{Amoretti:2016cad}
\bibitem{Amoretti:2016cad}
  A.~Amoretti, M.~Baggioli, N.~Magnoli and D.~Musso,
  %``Chasing the cuprates with dilatonic dyons,''
  JHEP {\bf 1606} (2016) 113
  doi:10.1007/JHEP06(2016)113
  [arXiv:1603.03029 [hep-th]].
  %%CITATION = doi:10.1007/JHEP06(2016)113;%%

  
  \bibitem{Bigazzi:2008zt} 
  F.~Bigazzi, A.~L.~Cotrone and A.~Paredes,
  ``Klebanov-Witten theory with massive dynamical flavors,''
  JHEP {\bf 0809}, 048 (2008)
  doi:10.1088/1126-6708/2008/09/048
  [arXiv:0807.0298 [hep-th]].
  
\bibitem{HoyosBadajoz:2008fw} 
  C.~Hoyos-Badajoz, C.~Nunez and I.~Papadimitriou,
  ``Comments on the String dual to N=1 SQCD,''
  Phys.\ Rev.\ D {\bf 78}, 086005 (2008)
  doi:10.1103/PhysRevD.78.086005
  [arXiv:0807.3039 [hep-th]].  
  
  
  \bibitem{Susskind:1998dq} 
  L.~Susskind and E.~Witten,
  ``The Holographic bound in anti-de Sitter space,''
  hep-th/9805114.
  
  
  \bibitem{Peet:1998wn} 
  A.~W.~Peet and J.~Polchinski,
  ``UV / IR relations in AdS dynamics,''
  Phys.\ Rev.\ D {\bf 59}, 065011 (1999)
  doi:10.1103/PhysRevD.59.065011
  [hep-th/9809022].
  
  %\cite{Girardello:1998pd}
\bibitem{Girardello:1998pd}
  L.~Girardello, M.~Petrini, M.~Porrati and A.~Zaffaroni,
  ``Novel local CFT and exact results on perturbations of N=4 superYang Mills from AdS dynamics,''
  JHEP {\bf 9812} (1998) 022
  doi:10.1088/1126-6708/1998/12/022
  [hep-th/9810126].
  %%CITATION = doi:10.1088/1126-6708/1998/12/022;%%
  
  %\cite{Freedman:1999gp}
\bibitem{Freedman:1999gp}
  D.~Z.~Freedman, S.~S.~Gubser, K.~Pilch and N.~P.~Warner,
  ``Renormalization group flows from holography supersymmetry and a c theorem,''
  Adv.\ Theor.\ Math.\ Phys.\  {\bf 3} (1999) 363
  [hep-th/9904017].
  %%CITATION = HEP-TH/9904017;%%
  
  
  %\cite{Bigazzi:2009gu}
\bibitem{Bigazzi:2009gu}
  F.~Bigazzi, A.~L.~Cotrone, A.~Paredes and A.~V.~Ramallo,
  ``Screening effects on meson masses from holography,''
  JHEP {\bf 0905} (2009) 034
  doi:10.1088/1126-6708/2009/05/034
  [arXiv:0903.4747 [hep-th]].
  %%CITATION = doi:10.1088/1126-6708/2009/05/034;%%
  


  %\cite{Nishioka:2009un}
\bibitem{Nishioka:2009un} 
  T.~Nishioka, S.~Ryu and T.~Takayanagi,
  ``Holographic Entanglement Entropy: An Overview,''
  J.\ Phys.\ A {\bf 42}, 504008 (2009)
  doi:10.1088/1751-8113/42/50/504008
  [arXiv:0905.0932 [hep-th]].
  
\bibitem{Klebanov:2007ws} 
  I.~R.~Klebanov, D.~Kutasov and A.~Murugan,
  ``Entanglement as a probe of confinement,''
  Nucl.\ Phys.\ B {\bf 796}, 274 (2008)
  doi:10.1016/j.nuclphysb.2007.12.017
  [arXiv:0709.2140 [hep-th]].
  
  \bibitem{Papadimitriou:2004rz} 
  I.~Papadimitriou and K.~Skenderis,
  ``Correlation functions in holographic RG flows,''
  JHEP {\bf 0410}, 075 (2004)
  doi:10.1088/1126-6708/2004/10/075
  [hep-th/0407071].
  
  
  %\cite{Batrachenko:2004fd}
\bibitem{Batrachenko:2004fd}
  A.~Batrachenko, J.~T.~Liu, R.~McNees, W.~A.~Sabra and W.~Y.~Wen,
  ``Black hole mass and Hamilton-Jacobi counterterms,''
  JHEP {\bf 0505} (2005) 034
  doi:10.1088/1126-6708/2005/05/034
  [hep-th/0408205].
  %%CITATION = doi:10.1088/1126-6708/2005/05/034;%%
  

  %\cite{Bigazzi:2009tc}
\bibitem{Bigazzi:2009tc}
  F.~Bigazzi, A.~L.~Cotrone and J.~Tarrio,
  ``Hydrodynamics of fundamental matter,''
  JHEP {\bf 1002} (2010) 083
  doi:10.1007/JHEP02(2010)083
  [arXiv:0912.3256 [hep-th]].
  %%CITATION = doi:10.1007/JHEP02(2010)083;%%

  
  \bibitem{Jones:2016iwx} 
  P.~A.~R.~Jones and M.~Taylor,
  ``Entanglement entropy in top-down models,''
  JHEP {\bf 1608}, 158 (2016)
  doi:10.1007/JHEP08(2016)158
  [arXiv:1602.04825 [hep-th]].
  
  

  
  \bibitem{Hoyos:2016cob} 
  C.~Hoyos, N.~Jokela, D.~Rodriguez Fernandez and A.~Vuorinen,
  ``Breaking the sound barrier in AdS/CFT,''
  arXiv:1609.03480 [hep-th].
  
\bibitem{Bedaque:2014sqa} 
  P.~Bedaque and A.~W.~Steiner,
  ``Sound velocity bound and neutron stars,''
  Phys.\ Rev.\ Lett.\  {\bf 114}, no. 3, 031103 (2015)
  doi:10.1103/PhysRevLett.114.031103
  [arXiv:1408.5116 [nucl-th]].
  
  \bibitem{Gockeler}
  M.~Gockeler, R.~Horsley, V.~Linke, P.~E.~L.~Rakow, G.~Schierholz and H.~Stuben,
  ``Is there a Landau pole problem in QED?,''
  Phys.\ Rev.\ Lett.\  {\bf 80} (1998) 4119
  doi:10.1103/PhysRevLett.80.4119
  [hep-th/9712244].
  %%CITATION = doi:10.1103/PhysRevLett.80.4119;%%
  
  \bibitem{Kim}
  S.~Kim, J.~B.~Kogut and M.~P.~Lombardo,
  ``On the triviality of textbook quantum electrodynamics,''
  Phys.\ Lett.\ B {\bf 502} (2001) 345
  doi:10.1016/S0370-2693(01)00201-5
  [hep-lat/0009029].
  %%CITATION = doi:10.1016/S0370-2693(01)00201-5;%%
  
  \bibitem{Gies}
  H.~Gies and J.~Jaeckel,
  ``Renormalization flow of QED,''
  Phys.\ Rev.\ Lett.\  {\bf 93} (2004) 110405
  doi:10.1103/PhysRevLett.93.110405
  [hep-ph/0405183].
  %%CITATION = doi:10.1103/PhysRevLett.93.110405;%%

  
  
\end{thebibliography}
\end{document}